\documentclass[11pt]{article}
\pdfoutput=1 
\usepackage{amssymb}
\usepackage{amsmath}
\usepackage{amstext}
\usepackage{mathrsfs}
\usepackage{graphicx,epsfig}
\usepackage{epsfig}
\usepackage{verbatim} 
\usepackage{fancybox}
\usepackage{color}
\usepackage{ulem}
\usepackage{enumitem}
\usepackage{bbm}
\usepackage{parskip}
\usepackage{dsfont}
\usepackage[numbers,sort&compress]{natbib}
\usepackage{framed}
\usepackage{anyfontsize}
\usepackage{braket}
\usepackage[separate-uncertainty=true,multi-part-units=single]{siunitx}

\usepackage{float}
\usepackage{subfig}

\usepackage{tikz}
\usetikzlibrary{decorations}
\pgfdeclaredecoration{complete sines}{initial}
{
    \state{initial}[
        width=+0pt,
        next state=upsine,
        persistent precomputation={\pgfmathsetmacro\matchinglength{
            \pgfdecoratedinputsegmentlength / int(\pgfdecoratedinputsegmentlength/\pgfdecorationsegmentlength)}
            \setlength{\pgfdecorationsegmentlength}{\matchinglength pt}
        }] {}
    \state{upsine}[width=\pgfdecorationsegmentlength,next state=downsine]{
        \pgfpathsine{\pgfpoint{0.25\pgfdecorationsegmentlength}{0.5\pgfdecorationsegmentamplitude}}
        \pgfpathcosine{\pgfpoint{0.25\pgfdecorationsegmentlength}{-0.5\pgfdecorationsegmentamplitude}}
    }
    \state{downsine}[width=\pgfdecorationsegmentlength,next state=upsine]{
        \pgfpathsine{\pgfpoint{0.25\pgfdecorationsegmentlength}{-0.5\pgfdecorationsegmentamplitude}}
        \pgfpathcosine{\pgfpoint{0.25\pgfdecorationsegmentlength}{0.5\pgfdecorationsegmentamplitude}}
}
    \state{final}{}
}

\newcommand{\Comment}[1]{{}}
\definecolor{darkblue}{rgb}{0.15,0.35,0.55}
\definecolor{comment}{rgb}{1,0.4,0.4}
\definecolor{reddish}{rgb}{0.65, 0.2, 0.2}
\usepackage[linktocpage=true]{hyperref}
\hypersetup{
colorlinks=true,
citecolor=darkblue,
linkcolor=reddish,
urlcolor=darkblue,
pdfauthor={},
pdftitle={},
pdfsubject={}
}

\newcommand{\be}{\begin{equation}}
\newcommand{\ee}{\end{equation}}
\newcommand{\bea}{\begin{eqnarray}}
\newcommand{\eea}{\end{eqnarray}}
\newcommand{\beas}{\begin{eqnarray*}}
\newcommand{\eeas}{\end{eqnarray*}}
\newcommand{\nn}{\nonumber}

\def\({\left(}
\def\){\right)}

\newcommand{\rd}{{\rm d}}

\newcommand{\Mpl}{M_{\rm Pl}}
\newcommand{\MBH}{M_{\rm BH}}
\newcommand{\JBH}{J_{\rm BH}}

\renewcommand{\Im}{\operatorname{Im}}
\renewcommand{\Re}{\operatorname{Re}}

\newcommand{\rs}{r_s}

\def\gsim{ \lower .75ex \hbox{$\sim$} \llap{\raise .27ex \hbox{$>$}} }
\def\lsim{ \lower .75ex \hbox{$\sim$} \llap{\raise .27ex \hbox{$<$}} }

\newcommand{\beq}{\begin{equation}}
\newcommand{\eeq}{\end{equation}}
\newcommand{\beqs}{\begin{equation*}}
\newcommand{\eeqs}{\end{equation*}}
\newcommand{\beqar}{\begin{eqnarray}}
\newcommand{\eeqar}{\end{eqnarray}}
\newcommand{\bal}{\begin{aligned}}
\newcommand{\eal}{\end{aligned}}

\def\dalam{\hbox
{\vrule\vbox{\hrule\hbox to 1ex{ \hfill}\kern 1 ex\hrule}\vrule}}

\def\1/2{\hbox{$ {1 \over 2}$ }}

\def\h{\hbar}
\def\i/h{{i \over \h}}

\def\xyma{\xymatrix@M.7em}
\def\xymas{\xymatrix@M.1em}
\newcommand{\ba}{\begin{eqnarray}}
\newcommand{\ea}{\end{eqnarray}}

\definecolor{darkred}{rgb}{0.7,0.3,0.3}
\definecolor{darkgreen}{rgb}{0.2,0.7,0.3}
\definecolor{greyish}{rgb}{.90,.90,.90}
\definecolor{greyish2}{rgb}{.96,.96,.96}
\definecolor{darkblue2}{rgb}{0.3,0.4,0.9}
\usepackage{pifont}
\usepackage{xcolor,colortbl}

\usepackage{tcolorbox}

\title{}
\author{}

\numberwithin{equation}{section}

\begin{document}
\setcounter{page}{1}
\renewcommand{\thefootnote}{\fnsymbol{footnote}}
~
\vspace{.80truecm}

\begin{center}
{\fontsize{24}{15} \bf Black hole superradiance\\\vskip 3pt with (dark) matter accretion}
\end{center}

\vspace{0.3cm}

\begin{center}
{\fontsize{13}{18}\selectfont
Lam Hui,${}^{\rm a}$\footnote{\href{mailto:lh399@columbia.edu}{\texttt{lh399@columbia.edu}}}
Y.T. Albert Law,${}^{\rm a, b}$\footnote{\href{mailto:ylaw1@g.harvard.edu}{\texttt{ylaw1@g.harvard.edu}}}
Luca Santoni,${}^{\rm c}$\footnote{\href{mailto:lsantoni@ictp.it}{\texttt{lsantoni@ictp.it}}}
Guanhao Sun,${}^{\rm a,d}$\footnote{\href{mailto:gs2896@columbia.edu}{\texttt{gs2896@columbia.edu}}}
\\[4.5pt]
Giovanni Maria Tomaselli,${}^{\rm e}$\footnote{\href{mailto:g.m.tomaselli@uva.nl}{\texttt{g.m.tomaselli@uva.nl}}}
Enrico Trincherini${}^{\rm f}$\footnote{\href{mailto:enrico.trincherini@sns.it}{\texttt{enrico.trincherini@sns.it}}}
}
\end{center}
\vspace{0.45cm}

 \centerline{{\it ${}^{\rm a}$Center for Theoretical Physics, Department of Physics,}}
 \centerline{{\it Columbia University, New York, NY 10027, USA}} 
 
  \vspace{.18cm}
 
 \centerline{{\it ${}^{\rm b}$Center for the Fundamental Laws of Nature,}}
 \centerline{{\it Harvard University, Cambridge, MA 02138, USA}}
 
  \vspace{.18cm}

\centerline{{\it ${}^{\rm c}$ICTP, International Centre for
    Theoretical Physics,}}
\centerline{{\it Strada Costiera 11, 34151, Trieste, Italy}}

	\vspace{.18cm}

\centerline{{\it ${}^{\rm d}$Department of Physics, University of California, San Diego, La Jolla, CA 92093, USA}}

  \vspace{.18cm}

\centerline{{\it ${}^{\rm e}$GRAPPA, Institute of Physics,}}
\centerline{{\it University of Amsterdam, Science Park 904, 1098 XH, Amsterdam, The
     Netherlands}}

  \vspace{.18cm}

\centerline{{\it ${}^{\rm f}$Scuola Normale Superiore, Piazza dei
    Cavalieri 7, 56126, Pisa, Italy and}}
\centerline{{\it INFN - Sezione di Pisa, 56100, Pisa, Italy}}

 \vspace{.1cm}

 \vspace{0.2cm}
\begin{abstract}
\noindent
Studies of black hole superradiance often focus on the growth of a cloud
in isolation, accompanied by the spin-down of the black hole. In this
paper, we consider the additional effect of the accretion of matter and
angular momentum from the environment. We show that, in many cases, the black hole evolves by drifting along the superradiance threshold, in which case the evolution of its parameters can be described analytically or
semi-analytically. We quantify the conditions under which accretion
can serve as a mechanism to increase the cloud-to-black hole mass
ratio, beyond the standard maximum of about $10 \%$. 
This occurs by a process we call over-superradiance, whereby accretion effectively
feeds the superradiance cloud, by way of the black hole. We give two explicit examples: accretion from
a vortex expected in wave dark matter and accretion from a baryonic
disk. In the former case, we estimate the accretion rate by using an analytical fit to the asymptotic behavior of the confluent Heun function. Level transition, whereby one cloud level grows while the other
shrinks, can be understood in a similar way.

\end{abstract}

\newpage

\setcounter{tocdepth}{2}
\tableofcontents

\vspace{1.0cm}

\renewcommand*{\thefootnote}{\arabic{footnote}}
\setcounter{footnote}{0}

\newpage

\section{Introduction}

The phenomenon of black hole superradiance has been known since the 1970s 
\cite{1972JETP351085Z,1973JETP3728S,Bardeen:1972fi,Press:1972zz,Damour:1976kh,
	Detweiler:1980uk}: mass and angular momentum can be extracted
from a Kerr black hole via an instability associated with the presence
of a light bosonic field. More recent work has emphasized axions or
axion-like-particles as particularly compelling examples of such a
field \cite{Arvanitaki:2009fg}, and explored observational signatures
such as the spin-down of black holes and gravitational wave emission
\cite{Arvanitaki:2010sy,Arvanitaki:2016qwi,Saha:2022hcd,Stott:2018opm,Davoudiasl:2019nlo}. 
See also \cite{Dolan:2007mj,Brito:2015oca} for state-of-the-art
computations and a comprehensive review.

\begin{figure}[t]
	\centering
	\includegraphics[width=0.75\textwidth]{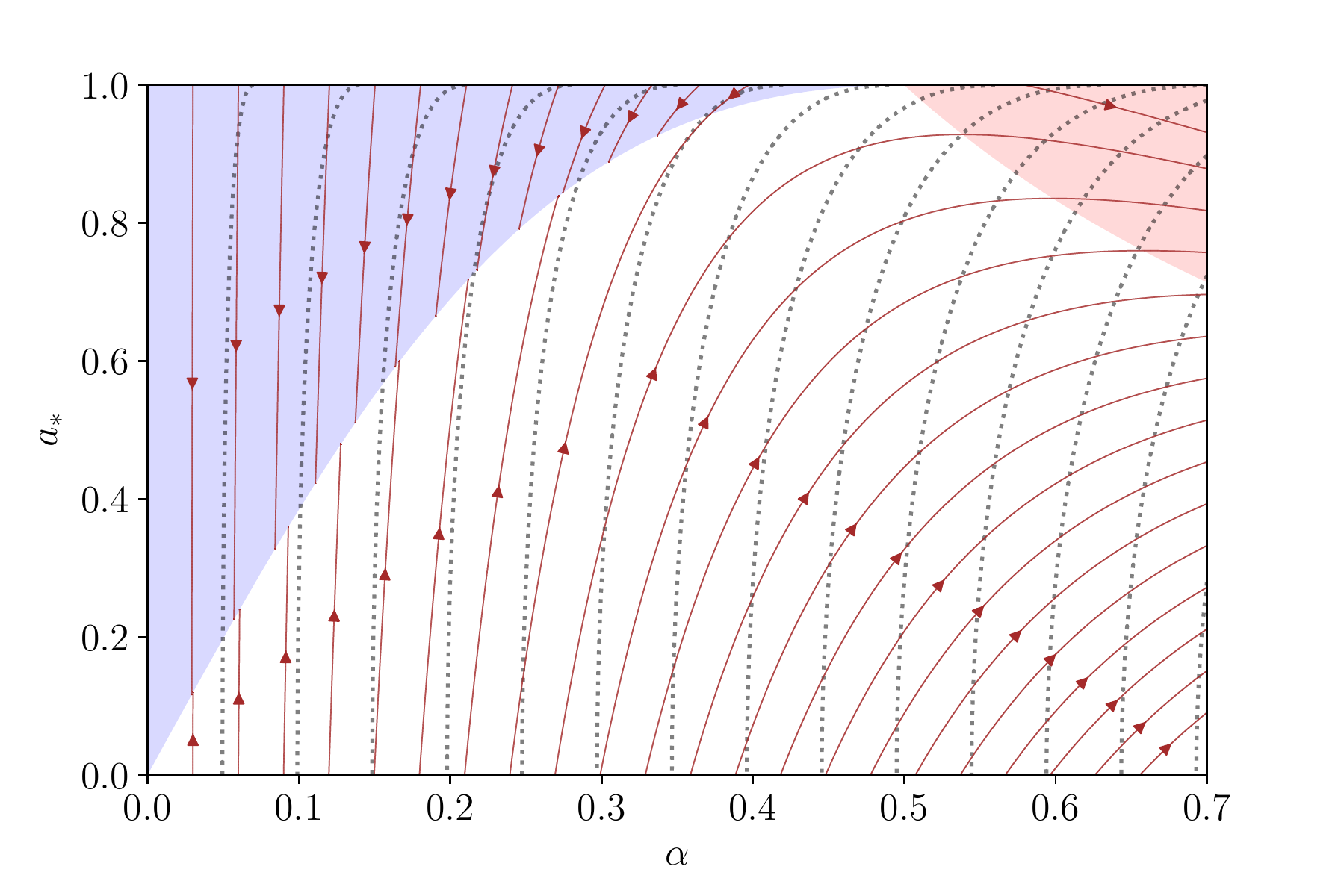}
	\caption{Trajectories of black hole evolution in the mass-spin
          (``Regge'') plane (as determined by
          \ref{eqn:mass-evolution-single-mode-approx} and
          \ref{eqn:J-evolution-single-mode-approx}). 
          Here, the black hole mass is encoded in
          $\alpha \equiv \mu r_s / 2$; the black hole spin is
          described by $a_* \equiv 2a/r_s$. The superradiance region
          is the upper-left, blue shaded region
          (\ref{eqn:super-inequality} or
          \ref{supercondition2}), assuming the scalar angular momentum
          quantum number $m=1$. 
          The red arrows in that region indicate the
          reduction of the black hole mass and spin as the
          $m=1$ superradiance cloud grows. The red arrows outside the blue
          region indicate the evolution of the black hole mass and
          spin, when the (non-superradiant) $m=1$ mode accretes 
          onto the black hole.
           In this case, the non-superradiant $m=1$ mode could be from 
           the cloud that was built up by superradiance (but now, the
           cloud shrinks
           and gives back mass to the black hole), or it could be from
          the ambient environment. The upper-right, shaded
          red region ($a_* > m/(2\alpha)$, from \ref{astardot}) indicates where
          the (non-superradiant) $m=1$ mode spins down the black hole
          (i.e., the black hole gains both mass and angular momentum
          from the mode, but gains mass faster such that $\dot a_* <
          0$). Lastly, the dotted gray lines are lines of constant
          horizon area; black hole evolution always respects the
          second law i.e.~horizon area increases.
          See Section \ref{sec:regge} for further discussions.
}
	\label{fig:regge-introductory}
\end{figure}

In this paper, we focus on the case of a scalar field; generalization to
higher spins is straightforward.
Consider a minimally coupled scalar $\Phi$ of mass $\mu$ on a Kerr background:
\begin{eqnarray}
	(- g^{\alpha\beta}\nabla_\alpha\nabla_\beta + \mu^2) \Phi = 0 \, .
\label{scalareqKG}
\end{eqnarray}
Imposing boundary conditions that $\Phi$ is (1)
ingoing at the horizon and (2) vanishes at infinity, it can
be shown that a solution of definite angular momentum numbers $\ell,
m$ has a discrete set of frequencies $\omega$, much like the hydrogen
atom. Superradiance refers to the possibility of energy and angular momentum extraction from the black hole, which occurs when the following inequality is satisfied
\begin{eqnarray}
	{\rm Re \,\omega} < {am \over r_s r_+} \, ,
	\label{eqn:super-inequality}
\end{eqnarray}
where $r_s = 2 GM_{\rm BH}$ is the Schwarzschild radius, $a$ is the
spin of the black hole (maximal spin is $a=r_s/2$), and $r_+$ is the
outer horizon. The combination $\Omega_+ \equiv a/(r_s r_+)$ is the angular velocity of the horizon. For the
hydrogen-like bound states, whenever (\ref{eqn:super-inequality}) is
satisfied, $\omega$ acquires a positive imaginary part
($\Phi \propto e^{-i\omega t}$), signalling an instability (see  Table~\ref{tab1} for a summary of the different cases). 
Typically, ${\rm Re \,\omega}$ is of the order of the
scalar mass $\mu$, and ${\rm Im\,\omega} \ll \mu$. 
The superradiance condition can thus be re-expressed as a condition on
the dimensionless spin of the black hole $a_* \equiv 2a/r_s$ as a
function of $\mu r_s/2$ (the gravitational radius to Compton scale
ratio). See the top-left shaded blue region of Figure~\ref{fig:regge-introductory} for an
illustration for $m=1$. This is the region of the parameter space in
which an initial scalar seed, no matter how small (even a quantum
fluctuation), can grow,
extracting both mass and angular momentum in the process, thereby
spinning down the black hole (see the red arrows in the blue
region). The upper bound on the mass of the cloud which grows in this way is, as we will see, about 10\% of the black hole mass \cite{Herdeiro:2021znw}.

The story we wish to tell starts when the black hole spins down to the
boundary of the superradiance region. In particular, let us remember
that black hole in nature rarely exists in isolation. The ambient matter, be it baryonic or dark matter,
can accrete onto the black hole. In most cases, as we will check
below, the accretion rate is small enough that the spin-down to the
superradiance boundary (the downward red arrows in the blue region)
is not significantly affected. But once the black hole approaches the
vicinity of the boundary, the mass and angular momentum extraction by
the cloud slows down considerably. Meanwhile, the ambient
accretion is still ongoing and can compete with superradiance. In
fact, to say ``compete'' does not quite convey the complete picture.
The two actually act in concert, in the following sense. The ambient accretion donates mass
and angular momentum to the black hole, while superradiance
extracts mass and angular momentum from the black hole at the same
time: accretion effectively
feeds the cloud---by way of the black hole.
The net result is a cloud that can grow to a significantly bigger size,
and a black hole that spins up and grows in mass. This behaviour has been seen for the first time through numerical computations in \cite{Brito:2014wla}, where baryonic accretion was considered. However, its dynamics has not received an explanation so far. The goal of this paper is to study this phenomenon analytically. Effectively, the
black hole climbs up the boundary of the superradiance region
(Figure~\ref{fig:regge-introductory}). In detail, the black hole
actually executes a trajectory ever so slightly above the
boundary---we thus call this \textit{over-superradiance-threshold-drift}, or 
\textit{over-superradiance} in short. It turns out an evolution 
slightly under the boundary is also possible,
where the cloud shrinks, giving mass back to the black hole.
We refer to this second way for the black hole to climb up the superradiance
boundary as \textit{under-superradiance-threshold-drift}, or
\textit{under-superradiance} in short.

We summarize here the main novel results of this work.
\begin{itemize}
\item We show that the threshold drift is a generic behaviour that onsets whenever the accretion timescale is much longer than the superradiance timescale. We do this in Section~\ref{sec:accretion+superradiance} with qualitative arguments, and in Appendix~\ref{sec:toy-model} quantitatively in a toy model, showing that the threshold drift is an attractor of the dynamics.
\item We formulate a precise condition to discriminate between over- and under-superradiance, see (\ref{eqn:under-over-def}). This inequality also tells whether the cloud's mass increases or decreases during the process.
\item We quantify the amount by which the trajectory deviates from the superradiance threshold. This is done expressing the change in the angular velocity of the horizon in terms of the mass of the cloud and the black hole parameters, see (\ref{eqn:deltaOmega}).
\item We consider the possibility that accretion occurs from the ambient dark matter. Building on earlier work
by \cite{Clough:2019jpm,Hui:2019aqm,Bamber:2020bpu}, we derive a widely applicable approximation for stationary accretion of scalar dark matter onto a black hole, see (\ref{eqn:guanhaos-fit}) and Appendix~\ref{app1}. This scenario is particularly appealing in the context of wave dark
matter, which is described by a light scalar field with a  mass
$\lesssim30 \text{ eV}$, exhibiting  wave phenomena (see, e.g.,
\cite{Hui:2016ltb,Niemeyer:2019aqm,Ferreira:2020fam,Hui:2021tkt} 
and references therein). This may or may not be the same scalar that leads to
superradiance. Wave dark matter naturally develops vortices due to wave
interference (\cite{Hui:2020hbq} and references therein), from
which the black hole can accrete mass and angular momentum in a way
that triggers over-superradiance.
\item We find analytical expressions to describe the threshold drift with dark matter accretion, determining the evolution of the cloud's mass as function of the black hole mass and the initial conditions, see equations (\ref{eqn:xc(x)})-(\ref{eqn:x_cx}). We find that dark matter accretion, albeit slow, can give a larger cloud-to-black hole mass ratio compared to baryonic accretion.
\item We describe the ``level transition'' happening between different states of the cloud when superradiance of multiple modes is considered. This was studied numerically in \cite{Ficarra:2018rfu}. Here, we show that this process can be understood as under-superradiance. The threshold drift phenomenon also happens during the level transisition, which can therefore be studied with the results derived earlier. We find analytically the coordinates of the relevant points of the trajectory in the Regge plane, see (\ref{eqn:rs''a''}).
\item We extend the results of \cite{Brito:2014wla} concerning baryonic accretion, providing a sharp bound on the maximum cloud-to-black hole mass ration attainable.
\end{itemize}

The outline of the paper is as follows.
We begin in Section \ref{sec:setup} with a 
brief review of superradiance in isolation 
(i.e., a single scalar mode present). The superradiance we are most
interested in is {\it bound} superradiance, which gives rise to the
growth of a cloud around the black hole, 
reducing the latter's mass and angular
momentum. In particular, the standard maximum
cloud-to-black-hole mass ratio of around $10\%$ is derived in
(\ref{eqn:limit-analytical}). 
We introduce the phenomenon of 
threshold
drift in Section \ref{sec:accretion+superradiance}, a process in
which the black hole moves along the superradiance threshold
in the mass-spin (``Regge'') plane, by combining the effects of
superradiance and accretion from the ambient environment.
The evolution of the black hole $+$ superradiance cloud system
is described by (\ref{eqn:threshold-derivative}) to
(\ref{eqn:over-under-superradiant}). The distinction between
over-superradiance and under-superradiance is explained here.
In Section \ref{sec:cases}, we present several different cases of
interest, for both accretion from (wave) dark matter (Section \ref{sec:dm-accretion}), and from a
baryonic accretion disk (Section \ref{sec:baryonic}). 
In Section \ref{sec:transition}, we demonstrate how the 
phenomenon of level transition, whereby one cloud level is 
depleted while another grows, can be understood in the same
threshold drift framework. We conclude in Section \ref{sec:discuss}
with a discussion of the observational implications and open
questions. A number of technical results can be found in the Appendices.

\paragraph{Notations and terminology.} We work in natural units, with $\hbar=c=1$. Newton's constant and the reduced Planck mass are related by $G=1/(8\pi\Mpl^2)$. Our metric signature will be $(-,+,+,+)$, with Greek letters standing for spacetime indices. The metric of a Kerr black hole of mass $\MBH$ and angular momentum $\JBH$ is
\begin{equation}
	\label{kerr}
	ds^2 = -\left(1 - {r_s r \over \varrho^2}\right)\rd t^2  - {2a r_s r {\,\rm sin}^2\theta
		\over \varrho^2}\rd t \rd \phi + {\varrho^2 \over \Delta} \rd r^2 + \varrho^2
	\rd \theta^2 + { (r^2 + a^2)^2 - a^2\Delta {\,\rm sin}^2\theta \over
		\varrho^2} {\,\rm sin}^2\theta \, \rd \phi^2 \, ,
\end{equation}
where $r_s \equiv 2 G M_{\rm BH}$ is the Schwarzschild radius, $a\equiv J_{\rm BH}/M_{\rm BH}$ is the spin parameter (which is taken to be nonnegative), $\varrho^2 \equiv r^2 + a^2 {\,\rm cos\,}^2\theta$ and $\Delta \equiv r^2 - rr_s + a^2$. The roots of $\Delta=0$ give the radii of the outer and inner horizons, $r_\pm \equiv r_s/2 \pm \sqrt{(r_s/2)^2 - a^2}$. The angular velocity of the outer horizon is $\Omega_+\equiv a/(\rs r_+)$ and the dimensionless black hole spin parameter is $a_* \equiv 2a/r_s$, ranging from 0 (non-rotating case) to 1 (extremal case).
We introduce the dimensionless quantity $\alpha\equiv\mu\rs/2$, where
$\mu$ is the scalar field mass.

A few words on terminology are in order. Accretion is
the process by which a black hole gains mass. This can occur via
accretion from the ambient environment (such as the surrounding dark
matter or baryonic disk), or accretion from the
cloud that was built up by superradiance (but is no longer in a
superradiant state due to the evolution of the black hole). 
Most of the time, by accretion, we implicitly
refer to the former, i.e.~ambient accretion.
We reserve the word ``cloud''
to describe the scalar cloud bound to the black hole, grown by superradiance.
We reserve the words ``ambient''
and ``environment'' to describe what is around the black hole other than
the superradiance cloud itself.

\section{Superradiance in isolation}
\label{sec:setup}

In this section, we set the stage by reviewing the massive
Klein-Gordon equation in Kerr background, describing solutions
involving fluxes of mass and angular momentum {\it into} and {\it out of} the
black hole. Superradiance refers to the latter possibility.
In particular, when the scalar is bound to the black hole (the scalar
vanishes far away from it), the superradiance is accompanied by an
instability: a scalar cloud grows around the black hole. 
The resulting backreaction on the black hole's mass and angular
momentum is described, and visualized in plots of the Regge plane.
Throughout this section, only a single mode (of angular momentum $m$)
is present.\footnote{Focusing on a single mode is a reasonable starting point,
  because as we will see, the timescales associated with different
  $m$'s are generally quite different.}
Most of the discussion focuses on this single mode being the
superradiant mode, though 
some of the discussion applies equally well if
the single mode refers to accretion from the ambient environment.
In the next section, we will study situations in which two modes are
present, including the most interesting case where one mode refers to
superradiance, and the other refers to ambient accretion.

\subsection{Fluxes and evolution equations}
\label{sec:fluxes-evolution}

Consider a scalar field $\Phi$ of mass $\mu$ in the Kerr background.
Throughout the paper we will ignore self-interactions of
$\Phi$.\footnote{See Section \ref{sec:discuss} for a discussion on when this is a good approximation.} Under this assumption,  the scalar obeys the  Klein-Gordon equation $(-g^{\alpha\beta}\nabla_\alpha\nabla_\beta + \mu^2) \Phi = 0$, which can be solved by decomposing it into a linear combination of
\begin{eqnarray}
\Phi_{\omega\ell m} = e^{-i\omega t} e^{im\phi} S_{\ell m}(\theta) R_{\omega\ell m}(r) \, ,
\end{eqnarray}
where $\omega$ is, in general, complex. Here, $e^{im\phi}S_{\ell m}(\theta)$ is a spheroidal harmonic, which reduces to the
spherical harmonic $Y_{\ell m}(\theta,\phi)$ if $a=0$ or $\omega = \mu$, and $R_{\omega\ell m}(r)$ is the
radial function, which depends on $\omega, \ell, m$ as well as the black
hole parameters $a$ and $r_s$ and scalar mass $\mu$. Both the spheroidal harmonic and the radial function are solution of the confluent Heun equation; details of the decomposed Klein-Gordon equation are given in Appendix \ref{app1}. The above expression is technically only valid if $\Phi$ is complex; if it were real, one should simply add the complex conjugate:
\begin{eqnarray}
\label{ccreal}
\Phi_{\omega\ell m} = e^{-i\omega t} e^{im\phi} S_{\ell m}(\theta) R_{\omega\ell m}(r) + \text{c.c.} \, .
\end{eqnarray}

Two central quantities of our study are the integrated energy and angular momentum fluxes across the horizon.
To derive them, we take the $(r,t)$ and $(r,\phi)$ components of the
scalar energy-momentum tensor in the Kerr background; here, for simplicity we consider
only one $(\omega,\ell,m)$ mode and drop the subscripts:\footnote{In the case of a real scalar field, the following
  expressions only hold in a time-averaged sense. More precisely,
the expressions for $T^\mu {}_\nu$ in terms of a real (and canonically
normalized) $\Phi$ has an extra
factor of $1/2$; this factor is canceled, once one expresses 
$\Phi$ as in Eq.~\eqref{ccreal} and evaluates the time-averaged
$T^\mu {}_\nu$.}
\begin{align}
T^r{}_t&=g^{rr}(\partial_r\Phi^*\partial_t\Phi+\partial_t\Phi^*\partial_r\Phi)=2\frac\Delta{\varrho^2}\Im(\omega R'^*R)|S|^2e^{2\Im(\omega)t},\\
T^r{}_\phi&=g^{rr}(\partial_r\Phi^*\partial_\phi\Phi+\partial_\phi\Phi^*\partial_r\Phi)=-2\frac\Delta{\varrho^2}m\Im(R'^*R)|S|^2e^{2\Im(\omega)t}.
\end{align}
From the near-horizon limit of the radial part of the Klein-Gordon equation,
\begin{equation}
\label{eqn:near-horizon-radial}
\Delta\frac{\rd}{\rd r}\biggl(\Delta\frac{\rd R}{\rd r}\biggr)+\rs^2r_+^2(\omega-m\Omega_+)^2R=0,
\end{equation}
we can extract the near-horizon behavior of $R(r)$,
\begin{equation}
R(r)\propto(r-r_+)^{-i\sigma},\qquad \sigma=\frac{\rs r_+(\omega-m\Omega_+)}{r_+-r_-},
\end{equation}
and evaluate the energy-momentum tensor at the horizon,
\begin{align}
\label{eqn:T^r_t}
T^r{}_t(r_+)&=2\frac{r_sr_+}{\varrho^2}(|\omega|^2-\Re(\omega)m\Omega_+)\Phi^*\Phi(r_+),\\
T^r{}_\phi(r_+)&=-2m\frac{r_sr_+}{\varrho^2}(\Re(\omega)-m\Omega_+)\Phi^*\Phi(r_+).
\label{eqn:T^r_phi}
\end{align}
The angular integrals of these quantities provide the total energy and momentum fluxes across the horizon, which we write as a variation of the mass and spin of the black hole:
\begin{align}
\label{eqn:mass-evolution-single-mode}
\dot M_{\rm BH}&=2\rs r_+(|\omega|^2-\Re(\omega)m\Omega_+)|R_+|^2,\\
\label{eqn:J-evolution-single-mode}
\dot J_{\rm BH}&=2\rs r_+m(\Re(\omega)-m\Omega_+)|R_+|^2,
\end{align}
where we have set $R_+\equiv R(r_+)$.

A few comments are in order about these expressions. First of all, they hold when only one $(\omega,\ell,m)$ mode is present. Because $T_{\mu\nu}$ is quadratic in the field, interference terms   appear when multiple modes are present. As we argue in Appendix \ref{app:superradiant-nonlinear}, these interference  terms are not expected to play a significant role in the cases we are interested in, because they oscillate much faster than the timescale of variation of $M_{\rm BH}$ and $J_{\rm BH}$.
Second, the frequency $\omega$ is determined by  
boundary
conditions. For a bound mode (i.e., $\Phi$ vanishes far away from the
black hole), $\omega$ turns out to be complex and discretized (see
\cite{Dolan:2007mj,Brito:2015oca} and Appendix \ref{app2}), 
while for an unbound mode (i.e., $\Phi$ does not vanish far away) 
$\omega$ is real and can
be interpreted as the energy of the scalar very far from the
black hole. In both cases, for applications of interest, $\omega\approx\mu$, so that the equations can be approximated as
\begin{align}
\label{eqn:mass-evolution-single-mode-approx}
\dot M_{\rm BH}&=2\rs r_+\mu(\mu-m\Omega_+)|R_+|^2,\\
\label{eqn:J-evolution-single-mode-approx}
\dot J_{\rm BH}&=2\rs r_+m(\mu-m\Omega_+)|R_+|^2.
\end{align}

Another important point is that, when we equate the fluxes to the
changes of the black hole parameters, we are no longer dealing with a
linear system. In other words, the Klein-Gordon equation, while
superficially linear in the scalar field, is not strictly so because the black
hole's geometry is modified by the scalar itself.
When the parameters of the black
hole change with time, the time dependence of the field will 
no longer
be strictly $e^{-i\omega t}$. 
Equations (\ref{eqn:mass-evolution-single-mode}) and
(\ref{eqn:J-evolution-single-mode}), or (\ref{eqn:mass-evolution-single-mode-approx}) and
(\ref{eqn:J-evolution-single-mode-approx}), therefore
need to be 
supplemented with information on the long-term 
evolution of the scalar field.

In case the mode under consideration is bound, this is 
usually done in a \textit{quasi-adiabatic} approximation
\cite{Brito:2014wla,Ficarra:2018rfu}, in which the growth
of the cloud is computed using the instantaneous value
of $\Im(\omega)$ (and we will further apply the
approximation $\Re(\omega)\approx\mu$). 
The total mass and angular momentum are kept constant.
The resulting evolution equations are
\begin{align}
\arraycolsep=16pt
\begin{array}{cc}
\dot M_{\rm BH}=-\dot M_c & \dot J_{\rm BH}=-\dot J_c\\
\dot M_c=2\Im(\omega)M_c & \dot J_c=\frac{m}\mu\dot M_c,
\end{array}
\label{eqn:nonlin-1-mode-super}
\end{align}
where $M_c$ and $J_c = m M_c/\mu$ are the mass and angular momentum carried by the
bound mode (the cloud). The role of $|R_+|$ in
Eqs.~(\ref{eqn:mass-evolution-single-mode-approx}) and
(\ref{eqn:J-evolution-single-mode-approx}) is thus replaced by 
$M_c$ in the system of equations here.

When multiple modes are present, an explicit model of the cloud profile is necessary to generalize the above equation \cite{Ficarra:2018rfu}.

If the mode under consideration is unbound, such as in the case of
scalar dark matter accretion from the environment, then 
$|R_+|$ in 
Eqs.~(\ref{eqn:mass-evolution-single-mode-approx}) and
(\ref{eqn:J-evolution-single-mode-approx}) would need to be 
connected to the scalar field value far away.
A stationary accretion flow solution 
(Appendix \ref{app1})
provides such a connection,
fixing the scalar amplitude at the horizon $|R_+|$
in terms of the scalar energy density $\rho$ far away.

As we will see, for our main conclusions only a few ingredients
of the equations above will be relevant, namely the ratio $\dot
J_c/\dot M_c$ (or $\dot J_{\rm acc}/\dot M_{\rm acc}$ for accretion
modes) and the fact that the mass in each individual bound state
$(n,\ell,m)$ grows as $\dot M_c/M_c\sim2\Im(\omega_{n\ell
  m})\sim(m\Omega_+-\mu)\alpha^{4\ell+5}$ (Appendix \ref{app2}).

\subsection{The Regge plane}
\label{sec:regge}

We are interested in the evolution of the black hole in the mass-spin
(``Regge'') plane (Figure \ref{fig:regge-introductory}). On the $x$-axis, we have $\alpha\equiv\mu\rs/2$, which is a measure of the mass of the black hole, while on the $y$-axis we have $a_* \equiv 2a/r_s$, i.e., the angular momentum to squared mass ratio.

Even though eventually we will be interested in situations where multiple modes are present, let us first gain some intuition on the Regge flow for a single mode. From (\ref{eqn:mass-evolution-single-mode})
and (\ref{eqn:J-evolution-single-mode}), we see that the scalar field extracts energy and angular momentum from the black hole when
\begin{equation}
\label{supercondition1}
\mu\approx\Re(\omega)<m\Omega_+.
\end{equation}
This is the superradiance condition.
If the mode of interest is bound (i.e., $\Phi$ vanishes far away),
the superradiance is accompanied by an instability, i.e.~${\,\rm Im\,}\omega > 0$, telling us that the scalar field
builds up around the black hole into a cloud.
In other words, the gravitational potential of the black hole does not
allow the superradiance generated scalar to escape, leading to a
run-away process.
It is useful to re-express the inequality as
\begin{eqnarray}
\label{supercondition2}
a_*  > { m / \alpha \over 1 + (m/2\alpha)^2}    \, ,\qquad\text{for}\quad\frac{2\alpha}{m}<1,
\end{eqnarray}
with $m$ understood to be positive. This region is shown in blue in Figure \ref{fig:regge-introductory}.

It is worth noting
that superradiance can occur with an unbound mode too.
If one were to remove the $\Phi \rightarrow 0$ boundary condition far away
from the black hole, $\omega$ can be real (and not discretized).
This means the extraction of mass and angular momentum from the black
hole does not occur by an exponential build up of the scalar cloud.
Rather, it occurs by sending the mass and angular momentum out to
infinity, a reverse accretion flow if you will. This is entry 3 in
Table \ref{tab1}, whereas bound superradiance is entry 1.

Note that the extraction of angular momentum implies, but is not implied by, $\dot a_*<0$, as
\begin{equation}
\label{astardot}
\dot a_*\propto\frac2{\rs^2}\bigl(m-2a_*\alpha\bigr)\biggl(\mu-\frac{ma_*}{2 r_+}\biggr),
\end{equation}
where we used $ \dot J_{\rm BH} =(m/\mu)\dot M_{\rm BH}$ and $\dot
M_{\rm BH}\propto(\mu-m\Omega_+)$. This expression also tells us
there is a region
$a_*>m/(2\alpha)$, 
not overlapping with (\ref{supercondition2}), where the black hole is
spun down even if its angular momentum increases. This region is shown
in red (top-right corner) in Figure \ref{fig:regge-introductory}.

The Regge trajectories in Figure \ref{fig:regge-introductory} are
obtained by computing $\dot a_* / \dot \alpha$ from 
Eqs.~(\ref{eqn:mass-evolution-single-mode-approx})
and (\ref{eqn:J-evolution-single-mode-approx}).
The red arrows in the blue region represent the Regge trajectories
when an $m=1$ (bound) superradiance cloud grow, 
reducing the black hole's mass
and angular momentum. The red arrows outside the blue region
show the Regge trajectories when a (non-superradiant) $m=1$ mode
accretes onto the black hole. Such a non-superradiant $m=1$ mode
could arise from the bound cloud that was previously built up by
superradiance (which now shrinks and gives back mass and angular
momentum to black hole), or it could be from the (unbound) 
ambient environment. 
Note that the field amplitude at the horizon $|R_+|$ gets scaled out
of the $\dot a_* / \dot \alpha$ ratio. However, the speed with which the black hole follows
these trajectories will depend on $|R_+|$. We will see below how
the timescale can be quite different for different scenarios.
See Table \ref{tab1} for a summary of the various scalar field
configurations of interest.

\begin{table}[tb]
\vspace{0.2cm}
\begin{center}
\begin{tabular}{@{}l|c|c}
\hline
{} & $\mu  < am/(r_s r_+)$ (superradiance) & $\mu  > am/(r_s r_+)$
\\ \hline
bound (complex $\omega$, ${\rm Re\,} \omega < \mu$) & 1. cloud grows
                                                      (${\rm Im\,}
                                                      \omega  > 0$)
                                                      \,\, \,&
                                                                      2. cloud
                                                                      shrinks (${\rm Im\,}
                                                      \omega  < 0$)
                                                               \, \\
{} & black hole shrinks \,\,\,\, \,\,\,& black hole grows \,\,\,\,\,\,\quad\quad
\\ \hline
unbound (real $\omega$, $\omega \ge \mu$) & 3. ambient mass/ang. mom.
                                           & 4. ambient
                                                              mass/ang. mom.
  \\
{} &  \,\,\quad \, extraction from black hole             &
                                                              \,\,\,\,
                                                         accretion
                                               onto
                                                              black hole
\\ \hline
\end{tabular}
\caption{Summary table for the different scalar configurations of
  interest around a black hole. The
  superradiance condition (\ref{eqn:super-inequality}) has been
  applied with the approximation $\Re\omega\approx\mu$.}
\label{tab1}
\end{center}
\end{table}

\subsection{Black hole spin down and growth of the superradiance cloud}
\label{sec:single-evolution}
 
Let us discuss in a bit more detail the case of bound
superradiance (entry 1 in Table \ref{tab1}). Solving (\ref{eqn:nonlin-1-mode-super}), when $\Im(\omega)$ is
properly expressed as a function of the mass and spin of the 
black
hole, gives the time evolution of the black hole parameters 
due to the
growth of the superradiance cloud. While such a solution
may not have an easy analytical expression, 
things simplify when the
time coordinate is factored out, i.e., when we only look at the
trajectory in the Regge plane. 
We will use a similar approach in
Section \ref{sec:accretion+superradiance} when putting 
accretion and
superradiance together, so let us describe how it works.

Because the extraction happens with a fixed angular momentum-to-mass ratio (and equal to $\dot J_c/\dot M_c=m/\mu$), (\ref{eqn:nonlin-1-mode-super}) implies
\begin{equation}
\frac{\rd}{\rd t}\biggl(M_{\rm BH}-\frac{\mu}mJ_{\rm BH}\biggr)=0.
\label{eqn:trajectory}
\end{equation}
This simple observation fully determines the trajectory 
followed by
the black hole in the Regge plane. Superradiance can only last until
$\Im(\omega)$ reaches zero (i.e., $\dot J_c=\dot M_c=0$), which means
that the black hole hits the threshold $\mu=m\Omega_+$, see Figure
\ref{fig:regge-introductory}. As long as no other states are
considered, this will be a point of stable equilibrium for the system:
moving above the threshold in the Regge plane will cause the cloud to
become superradiant, pushing the black hole down again; moving below
the threshold will cause the cloud to decay, giving mass and angular
momentum to the black hole, pushing the black hole back to the
threshold. Intersecting the trajectory (\ref{eqn:trajectory}) with the threshold $\mu=m\Omega_+$, we can find analytically the final parameters $(\rs',a')$ of the black hole in terms of the initial ones $(\rs,a)$:
\begin{equation}
\label{eqn:analytic-formulae}
\frac{\mu\rs'}m=\frac{1-\sqrt{1-(2(\mu\rs/m)(1-\mu a/m))^2}}{2(\mu\rs/m)(1-\mu a/m)},\qquad \frac{a'}{\rs'}=\frac{\mu\rs}m\Bigl(1-\frac{\mu a}m\Bigr).
\end{equation}
The cloud mass at the end of the process will be
$M_c=(\rs-\rs')/(2G)$. The maximum ratio between the mass of the cloud
and the mass of the black hole achievable with the evolution of a
single state can be thus obtained from the formula above:\footnote{This estimate is precisely equivalent to the one presented in \cite{Herdeiro:2021znw}, where the authors compute instead $\max\bigl\{(\rs-\rs')/\rs\bigr\}=9.73\%$.}
\begin{equation}
\max\Bigl\{\frac{M_c}{M_{\mathrm{BH}}}\Bigr\}=\max\Bigl\{\frac{\rs-\rs'}{\rs'}\Bigr\}=10.78\%,\qquad\text{for}\quad\frac{\mu\rs}m\approx0.24\quad\text{and}\quad\frac{a}\rs=0.5.
\label{eqn:limit-analytical}
\end{equation}

How much time does it take to grow the cloud? Although the system of equations (\ref{eqn:nonlin-1-mode-super}) is nonlinear, the nonlinearities are negligible as long as the size of the cloud is small enough to not make $\Im(\omega)$ change appreciably. As a consequence, for a cloud growing from a small seed, for example a quantum fluctuation, we can estimate the growth time as
\beq
T_{\rm growth}\approx\frac{\log(M_{c}/M_{c{\rm,\,seed}})}{2\Im(\omega(\rs,a))},
\label{eqn:t-growth}
\eeq
where $\rs$ and $a$ are the initial black hole parameters. For a quantum fluctuation, we have $M_{\rm c, seed}r_{c}\sim1/2$, where $r_{c}=\rs n^2/(2\alpha^2)$ and $n$ is the principal quantum number of the cloud, giving
\beq
\log\biggl(\frac{M_{c}}{M_{\rm c, seed}}\biggr)\sim175.5+\log\biggl(\frac{M_{c}}{M_{\rm BH}}\frac{M_{\rm BH}^2}{M_\odot^2}\frac{n^2}{\alpha^2}\biggr).
\eeq
The growth rate $\Im(\omega(\rs,a))/\mu$ varies by orders of magnitude
across the instability region of a given mode, reaching a maximum of
about $10^{-7}$ for $\alpha=0.42$, $m=1$ \cite{Dolan:2007mj}. $T_{\rm
  growth}$ can then be as short as several hours for stellar black
holes, and $10^5$--$10^6$ years for supermassive black holes.

\section{Threshold drift: combining superradiance with accretion}
\label{sec:accretion+superradiance}

In this section we explain in detail the evolution of a black hole in
the presence of both a cloud from superradiance {\it and} accretion
from the ambient environment.\footnote{The accretion could in principle also be from {excited states of the cloud which still undergo superradiance, thus falling in the first case of Table~\ref{tab1}, providing a ``negative accretion'', or extraction of mass. This is relevant for level 
transition that will be discussed in Section \ref{sec:transition}. 
The threshold drift discussion in
the current section applies equally well, for accretion from the ambient
environment, as for accretion from the cloud.}
}
In Section \ref{sec:setup} we showed that the evolution
of a black hole and its superradiance cloud is determined by the
nonlinear\footnote{The nonlinearity refers to the implicit dependence
  of $\Im(\omega)$ on the mass and spin of the black hole.} equations
(\ref{eqn:nonlin-1-mode-super}). It is worth stressing that these equations neglect potentially
important effects, like the depletion of the cloud due to
gravitational waves, or self-interactions of the scalar field which
mix different levels. We will briefly discuss them in Section \ref{sec:discuss}.

In Section \ref{sec:single-evolution} we described the evolution of a
black hole from its initial ``starting point'', to its final
meta-stable, or at least long-lived, state, surrounded by a boson
cloud of a single $m$ mode. Processes involving either additional states or external
effects are needed to drive the gravitational atom away from this final
position in the Regge plane. In this section, we take the endpoint of
the single-mode evolution as an initial condition, and focus on the
case where the system is fed mass and angular momentum from the
outside.\footnote{It is also possible, as we will see in Section
  \ref{sec:cases}, that the black hole reaches the threshold ``from
  below'', instead of from above, for example because of the same
  accretion mechanism that drives its subsequent evolution. The way
  the black hole arrives at the threshold does not matter for the
  discussion here.} How does the gravitational atom respond?

Suppose that some external fluxes of mass and of angular momentum change the parameters of the black hole as
\begin{equation}
2G\dot M_{\rm acc}=\frac{\rd\rs}{\rd t}\bigg|_{\mathrm{acc}},\qquad2G\dot J_{\rm acc}=\frac{\rd(a\rs)}{\rd t}\bigg|_{\mathrm{acc}}.
\end{equation}
where the label ``acc'' stands for accretion. 
Like in the last section, we assume the bound, superradiance cloud is
described by a single $(n,\ell,m)$ state. 
The evolution equations (\ref{eqn:nonlin-1-mode-super}) get modified
by accretion as follows:
\begin{align}
\label{eqn:M:acc+superr}
\dot M_{\rm BH}&=\dot M_{\mathrm{acc}}-\dot M_c,\\
\label{eqn:J:acc+superr}
\dot J_{\rm BH}&=\dot J_{\rm acc}-\frac{m}\mu\dot M_c,\\
\label{eqn:R-Im-omega}
\dot M_c&=2\Im(\omega)M_c,
\end{align}
where we approximated $\omega\approx\mu$. At the end of the previous
section, we have seen that the superradiance timescale can be very
short compared to other astrophysical processes such as accretion (see
Eq.~\ref{eqn:t-growth}), going from hours to $10^5$--$10^6$ years
depending on the mass of the black hole. 
Thus, superradiance, when it is operative, will tend to move the black
hole to the superradiance threshold in the Regge plane where it is
turned off. And it is when the black hole is very close to the threshold
that accretion can compete with superradiance.
The crucial observation for
our discussion is that, whenever the superradiance timescale is much
shorter than the accretion timescale (assumed henceforth, which we
will verify later), the system described by
Eqs.~(\ref{eqn:M:acc+superr}), (\ref{eqn:J:acc+superr}) and
(\ref{eqn:R-Im-omega}) will closely follow the superradiance threshold
line defined by $\mu=m\Omega_+$ during its accretion-driven evolution
in the Regge plane, as long as the cloud has a high enough occupation
number. We can see how this works in a few different ways.

\begin{figure}[t]
	\centering
	\includegraphics[width=\textwidth]{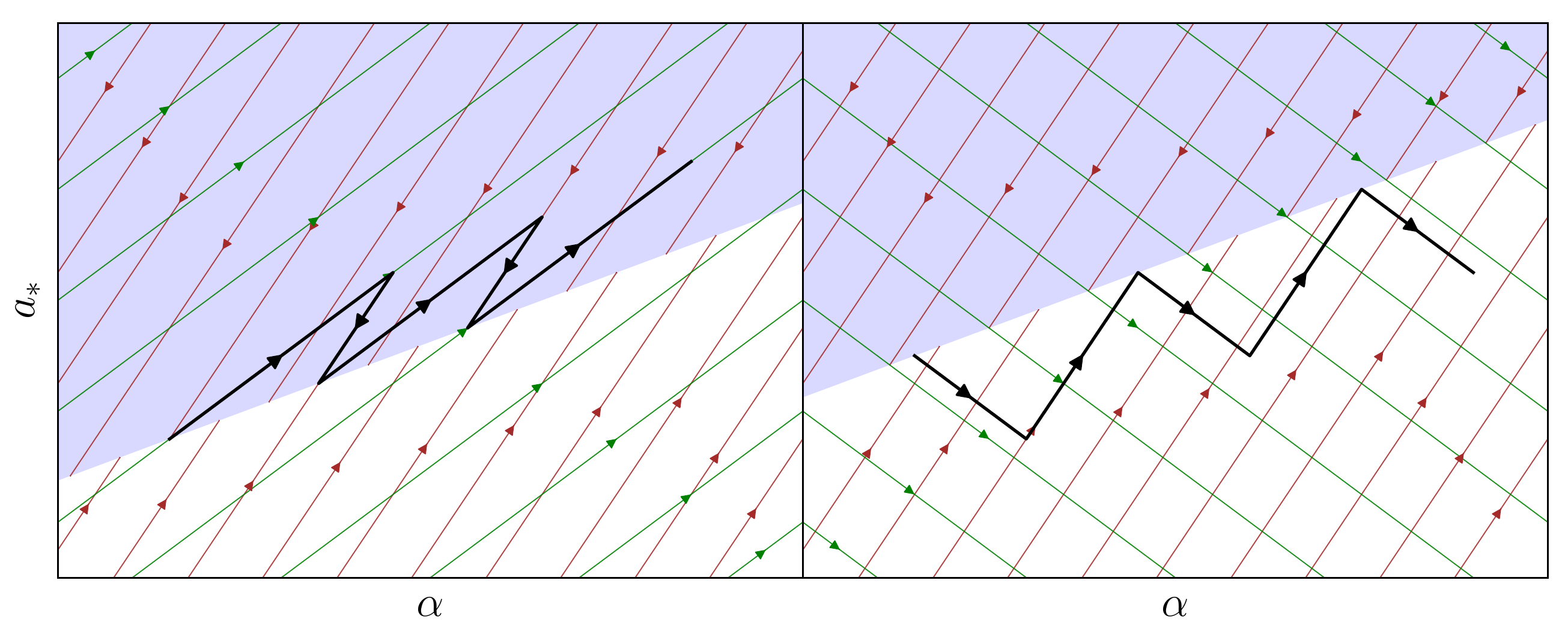}
	\caption{Pictorial (i.e., not literal) representation of
          threshold drift, in the cases of
          over- (left) and under-superradiance (right),
          defined by (\ref{eqn:under-over-def}). The image is
          schematic and represents a zoom-in of the
          Regge plane near the superradiance threshold (see Figure
          \ref{fig:regge-introductory} for a zoom-out). The shaded
          blue area denotes the region of superradiance
          instability. The red lines in the blue region show the
          trajectories of the black hole under superradiance: 
          the superradiance cloud
          grows and the black hole shrinks. The red lines outside the blue region
          show the opposite: suppose a cloud grown by superradiance already
          exists around the black hole, these red lines depict the
          evolution of the black hole as the cloud shrinks and gives
          back mass and angular momentum to the black hole. 
          In other words, the red lines describe the effects of the
          cloud terms in \eqref{eqn:M:acc+superr} and \eqref{eqn:J:acc+superr}.
          The green lines depict how the black hole would evolve under
          accretion alone (accretion terms in \eqref{eqn:M:acc+superr} and
\eqref{eqn:J:acc+superr}). The left/right panel depicts the case where 
          accretion tends to move the black hole above/below the
          superradiance threshold. The black zig-zag line is a pictorial representation of the trajectory followed by the black hole, which in reality is a smooth drift to the right, either slightly above (left) or below (right) the threshold.}
	\label{fig:zig-zag}
\end{figure}

Let us start with  a simple intuitive explanation. With reference to
Figure \ref{fig:zig-zag}, let us consider a black hole sitting exactly
on the superradiance threshold, $\mu=m\Omega_+$.  The second terms on
the right-hand sides of Eqs.~\eqref{eqn:M:acc+superr} and
\eqref{eqn:J:acc+superr} vanish, and we can think of accretion as a
process that  tends to drive the black hole slightly away from the
threshold. During this first step of the evolution, depending on the slope of accretion
with respect to the threshold, the black hole will end up either above
or below it. 

Consider the case where the black hole is driven
slightly above the threshold (left panel), superradiance then kicks
in, and because it is very efficient (unless one is on the threshold),
quickly moves the black hole back to the threshold. During this second
step, the superradiance cloud grows further in mass.
The black hole loses mass in this second step, but
the combined action of the first (accretion) and second
(superradiance) steps is such that the black hole has a net gain in
mass. The two-step combination repeats itself, and 
as a result, the black hole climbs up along the
superradiance threshold. (It can never climb down, by virtue of the
second law of thermodynamics; see Figure \ref{fig:regge-introductory}.) We refer to this phenomenon as
over-superradiance-threshold-drift, or over-superradiance in short.

Conversely, consider the case where the first (accretion) step takes
the black hole below the threshold (right panel). Remember the superradiance cloud
is still there, and because the black hole is below threshold, the
cloud will in fact shrink and give mass back to the black hole. 
This second step involves the existing superradiance cloud, but in a
non-superradiant state, i.e.~${\,\rm Im\,}\omega < 0$, as opposed to
${\,\rm Im\,}\omega > 0$, implying the scalar field decreases in value,
that is to say, cloud loses mass. The net effect of the first and second steps
is once again to increase the black hole's mass, and as a result, the black hole
climbs up along the superradiance threshold.
We refer to this phenomenon as
under-superradiance-threshold-drift, or under-superradiance in short.

Of course, this discrete ``zig-zag'' description of over- or under-superradiance
should not be taken literally.
In reality, the black hole's evolution in the Regge plane is smooth:
it drifts along a trajectory that closely hugs the superradiance
threshold, where the effects of accretion and superradiance (or cloud
decay) finely complement each other. We call this phenomenon threshold drift.
It is worth emphasizing again that the black hole can only drift to
the right, in the direction of increasing its mass. This is because
the other direction is forbidden by the second law of the black hole
thermodynamics, as the area of the event horizon would reduce (see
Appendix \ref{sec:superradiance-area} and Figure
\ref{fig:regge-introductory}). The superradiance 
trajectories at precisely the threshold are in fact parallel to lines of
constant area; the accretion trajectories must intersect them
and point to the right.

This threshold drift phenomenon can be seen in numerical solutions for
the evolution of the black hole in the Regge plane. For example, this
explains why  in \cite{Brito:2014wla}, where accretion from a baryonic
disk was
taken into account,  the numerical evolution tracked the superradiance
threshold for
a significant part of the black hole's history. In Section
\ref{sec:cases}, we will examine the case of baryonic accretion more
closely, and present a semi-analytic way to understand the black
hole's evolution.

We note that it is possible to understand the dynamics of the system in the Regge plane
in terms of a simpler toy model. This is described in detail in  Appendix \ref{sec:toy-model}, where we show that the threshold drift is an attractor of the dynamics as long as the mass of the cloud has a sizable value.

Having established that the system follows a trajectory that lies very
close to the superradiance threshold in the Regge plane, we can thus
enforce the condition $\mu=m\Omega_+$ in the equations
\eqref{eqn:M:acc+superr} and \eqref{eqn:J:acc+superr}, reducing it to
a single ordinary differential equation. Let us first take
the time derivative of $\mu=m\Omega_+$ (or equivalent, $a_* =
(m/\alpha)/(1 + (m/2\alpha)^2)$; see \ref{supercondition2}):
\begin{equation}
\frac\mu{m}\frac{\rd(a\rs)}{\rd t}=\frac{x^4+3x^2}{(1+x^2)^2}\frac{\rd\rs}{\rd t},
\label{eqn:threshold-derivative}
\end{equation}
where we defined $x\equiv\mu\rs/m$. Note that the superradiance
instability region, as well as its threshold, spans the region
$0<x<1$. Combining \eqref{eqn:M:acc+superr} and
\eqref{eqn:J:acc+superr} to eliminate the dependence on the mass of
the cloud, and then plugging (\ref{eqn:threshold-derivative}) in, we
obtain an equation for the evolution of the black hole's
mass:
\begin{equation}
\frac{1-x^2}{(1+x^2)^2}\frac{\rd x}{\rd
  t}=\bigg(1-\frac\mu{m}\frac{\dot J_{\mathrm{acc}}}{\dot
  M_{\mathrm{acc}}}\biggr)\frac{\mu}m \frac{\rd\rs}{\rd
  t}\bigg|_{\mathrm{acc}} ,
\label{eqn:accretion-effective-x-evolution}
\end{equation}
where $\rd r_s/\rd t |_{\rm acc} \equiv 2 G \dot M_{\rm acc}$ can be
thought of as $\dot r_s$ due to accretion alone.
For given rates of mass and angular momentum accretion, equation
\eqref{eqn:accretion-effective-x-evolution} describes how the mass of
the black hole, $M_{\mathrm{BH}} \equiv mx/(2G\mu)$, evolves in time, as long
as the superradiance cloud with azimuthal number $m$ has high enough
mass to keep the black hole pinned at the superradiance threshold. 
While this process takes place, the cloud's mass $M_c \equiv mx_c/(2G\mu)$ evolves according to
\begin{equation}
\frac{\rd(x+x_c)}{\rd t}=\frac{\mu}m
\frac{\rd\rs}{\rd t}\bigg|_{\mathrm{acc}} 
\label{eqn:accretion-effective-xc-evolution}
\end{equation}
due to conservation of mass. Beware that if and when the cloud loses enough mass (say, at $x_c=0$), it will no longer be able to keep the black hole near the threshold, and this effective description of the evolution will break down. Equations \eqref{eqn:accretion-effective-x-evolution} and \eqref{eqn:accretion-effective-xc-evolution} can be combined to give
\begin{equation}
\frac{\rd x_c}{\rd t}=\biggl(\frac{1}{1-\mu\dot J_{\mathrm{acc}}/(m\dot M_{\mathrm{acc}})}\frac{1-x^2}{(1+x^2)^2}-1\biggr)\frac{\rd x}{\rd t}.
\label{eqn:over-under-superradiant}
\end{equation}
Looking at the sign of the parenthesis on the right-hand side, this
equation makes it easy to tell whether we have
\begin{equation}
\begin{aligned}
\text{over-superradiance}\quad&\longleftrightarrow\quad \frac{1}{1-\mu\dot J_{\mathrm{acc}}/(m\dot M_{\mathrm{acc}})}\frac{1-x^2}{(1+x^2)^2}>1\\
\text{under-superradiance}\quad&\longleftrightarrow\quad
\frac{1}{1-\mu\dot J_{\mathrm{acc}}/(m\dot
  M_{\mathrm{acc}})}\frac{1-x^2}{(1+x^2)^2}<1 \, ,
\end{aligned}
\label{eqn:under-over-def}
\end{equation}
where once again, $x$ is defined as $\mu r_s / m$. 
Note that, due to the shape of the function $(1-x^2)/(1+x^2)^2$, any
given fixed ratio $\dot J_{\mathrm{acc}}/\dot
M_{\mathrm{acc}}$ satisfying $0<\dot J_{\mathrm{acc}}/\dot
M_{\mathrm{acc}}<m/\mu$ 
will produce over-superradiance for a
sufficiently small $x$ (small black hole mass), but it will eventually
turn into under-superradiance as $x$ approaches 1.

The presence of a superradiance cloud thus acts as a \textit{glue}
that keeps the black hole attached to the superradiance threshold. The
glue gets enhanced by over-superradiance (because the cloud grows),
and weakened by under-superradiance (because the cloud shrinks). 
If the cloud is completely dissipated, either by under-superradiance or some other process, the threshold drift ends, with the black hole moving away from it, following the Regge trajectories determined by accretion.

We can quantify the strength of this glue by computing the amount by which the actual trajectory of the black hole deviates from the threshold. This can be done by combining (\ref{eqn:accretion-effective-x-evolution}) and (\ref{eqn:over-under-superradiant}) as
\beq
\dot M_c=\frac1{1-x^2}\biggl((1+x^2)^2\frac\mu{m}\dot J_{\mathrm{acc}}-(x^4+3x^2)\dot M_{\mathrm{acc}}\biggr)
\eeq
and then using $\dot M_c=2\Im(\omega)M_c$. Because $\Im(\omega)$ depends on the black hole's mass and spin, one can extract the deviation of the threshold, which we can quantify as a small change in the angular velocity of the horizon: $\Omega_+=\mu/m+\delta\Omega_+$, with $\delta\Omega_+\ll\Omega_+$. In detail, using the result from \cite{Baumann:2019eav}, we evaluate $\Im(\omega)$ near the threshold to linear order as
\beq
\Im(\omega)\big|_{m\Omega_+\approx\mu}\approx\mathcal I\,\delta\Omega_+,\qquad \mathcal I=\frac{m^{4\ell+6}(n+\ell)!}{4n^{2\ell+4}(n-\ell-1)!}\frac{\ell!^4}{(2\ell)!^2(2\ell+1)!^2}\biggl(\frac{1-x^2}{1+x^2}\biggr)^{2\ell} \frac{x^{4\ell+5}}{1+x^2},
\eeq
so that
\beq
\delta\Omega_+=\frac1{2\mathcal IM_c}\biggl(\frac{(1+x^2)^2}{1-x^2}\frac\mu{m}\dot J_{\mathrm{acc}}-\frac{x^4+3x^2}{1-x^2}\dot M_{\mathrm{acc}}\biggr).
\label{eqn:deltaOmega}
\eeq
The inverse dependence on $M_c$ shows, as anticipated, that a larger cloud keeps the black hole closer to the threshold. It is also easy to confirm that $\delta\Omega_+\gtrless0$ when the over-/under-superradiance condition (\ref{eqn:under-over-def}) is satisfied.

So far, we have neglected any other effect that is able to change to
the total mass of the system. While this may be a good approximation
if $\Phi$ is a complex scalar field, a real field undergoes an
inevitable decay via the emission of gravitational waves \cite{Yoshino:2013ofa}. This
adds an extra source term to the right-hand side of
(\ref{eqn:accretion-effective-xc-evolution}). Such an extra source of
cloud mass loss does not change the previous criterion to distinguish
between over-/under- superradiance, nor the time evolution of the black hole parameters during the threshold drift, as Eq.~(\ref{eqn:accretion-effective-x-evolution}) stays unchanged.
However, it does change the duration of the threshold drift, as the
cloud will disappear faster. Moreover, this mechanism for cloud mass
loss becomes more important as the black hole gains mass during
threshold drift, since the radiated power goes as $\alpha^{4\ell+10}$.

\section{Examples and cases of interest}
\label{sec:cases}

In this section, we apply the approach developed in Section
\ref{sec:accretion+superradiance} to three physical scenarios. First,
in Section \ref{sec:dm-accretion}
we consider cases where the black hole accretes dark matter from
the ambient environment, and the dark matter is itself a scalar field,
which might or might not be the same as the scalar making up the
superradiance cloud. We will show in particular (see ``Case 2'' below)
that, for 
certain choices of the parameters, it is possible to grow a cloud of
mass up to 
roughly a third of the black hole mass, well beyond the standard 
$\sim$10\% discussed in Section \ref{sec:setup}.   
Then, in Section \ref{sec:transition} we consider the phenomenon of 
\textit{level transition} using the language of under-superradiance.
Lastly, in Section \ref{sec:baryonic}
we study the case where the ambient accretion is sourced by
a baryonic disk. The same phenomena of over- and
under-superradiance occur here, with the advantage that disk
accretion can be more efficient, leading to a substantial cloud build-up
in a shorter amount of time.

\subsection{(Wave) dark matter accretion}
\label{sec:dm-accretion}

Black holes in nature are inevitably surrounded by dark matter.
If the dark matter is comprised of a scalar field, much of our earlier
discussion regarding the mass and angular momentum fluxes into the
black hole horizon applies to dark matter as well. 
A concrete, compelling example is the axion, or axion-like-particles 
(see \cite{Marsh:2015xka,Graham:2015ouw} for reviews). We assume their self-interaction strength is
sufficiently weak to be ignored, but will return to a discussion of
this in Section \ref{sec:discuss}. The same axion could be both the
dark matter of the ambient environment, as well as the scalar that
makes up the superradiance cloud. Or the two could be different scalar
fields. We will use $\mu', m'$ to refer to the mass and angular
momentum of the dark matter scalar, and $\mu, m$ to refer to
the mass and angular momentum of the superradiance scalar.
In the language of Table \ref{tab1}, the dark matter scalar from
the ambient environment is ``unbound'' whereas the scalar of the
superradiance cloud is ``bound''. 

To determine the dark matter ambient accretion rate onto the
black hole, we use Eqs.~(\ref{eqn:mass-evolution-single-mode-approx})
and (\ref{eqn:J-evolution-single-mode-approx}), with the horizon
scalar amplitude $R_+ \equiv R_{\rm acc} (r_+)$ fixed by using the stationary accretion flow
solution which connects it to the scalar amplitude far away $R_{\rm
  acc} (r_i)$, at a radius we call $r_i \gg r_+$. This is described in detail in Appendix
\ref{app1}, giving us the following useful approximations, from wave
to particle limits:\footnote{Approximate fitting formulae are needed if one wants to obtain an approximate analytical treatment and not to solely rely on a numerical study. This is due to the nature of the scalar wave equation \eqref{scalareqKG}, which is of the (confluent) Heun type (see Appendix~\ref{app1} for further details---see also Refs.~\cite{Bonelli:2021uvf,Bonelli:2022ten} for recent progress on connection formulae for the Heun function).}
\begin{equation}
	\frac{|R_{\rm acc}(r_+)|^2}{|R_{\rm acc}(r_i)|^2}=\begin{cases}
		\biggl(\dfrac{r_i}{\rs}\biggr)^{3/2} & \quad   \qquad \dfrac12(\ell'+1)\lesssim\mu'\rs \qquad \qquad  \text{(Particle)}\; ,\\
		\biggl(\dfrac{r_i}{\rs}\biggr)^{3/2}\biggl(\dfrac{2\mu'\rs}{\ell'+1}\biggr)^{6\ell'+3} & \quad    2\sqrt{\dfrac{\rs}{r_i}}\ll\mu'\rs\lesssim\dfrac12(\ell'+1) \quad   \text{(Intermediate)}\; ,\\
		\biggl(\dfrac{r_i}{\rs}\biggr)^{-2\ell'} &  \quad  \qquad \mu'\rs\ll2\sqrt{\dfrac{\rs}{r_i}} \qquad    \quad \; \text{(Wave/Ultralight)}\; ,
	\end{cases}
	\label{eqn:guanhaos-fit}
\end{equation}
where $\ell'$ is the second quantum number of the accreting mode and
$\mu'$ is the mass of the field, which we label differently from $\mu$
for the sake of generality, to allow the dark matter field and the
superradiance field to be different. The precise expressions and
bounds in \eqref{eqn:guanhaos-fit} actually depend on the
dimensionless spin $a_*$ and magnetic quantum number $m'$; however, as
shown in Appendix~\ref{app1}, their effects on the estimates
\eqref{eqn:guanhaos-fit} are within $O(1)$ unless $a_*>0.95$. In other
words, the spin of the black hole does not have a significant impact
unless it is close to extremal. These
expressions generalize to non-zero angular
momentum those given in \cite{Hui:2019aqm}.

The quantity $r_i$ is taken to be the radius at which the dark matter
density matches the typical density $\rho_i$ in the broader
environment, i.e.~$\rho_i=\rho(r_i)$. A
quantitative estimate of $r_i$ is needed to fix the amplitude of the
accreting mode, and thus the timescale of the whole process.
For spherically symmetric accretion, we follow \cite{Hui:2019aqm}
and take $r_i$ to be the radius of impact of the black
hole (the radius at which the gravitational potential of the black
hole is similar to that of the dark matter halo), i.e.~$r_i/r_s \sim 10^6 (v_{\rm typical}/300 {\,\rm km/s})^{-2}$ where
$v_{\rm typical}$ is the velocity dispersion of the dark matter halo.
For dark matter accretion flow of angular momentum (per particle)
$m' \ne 0$, we will take $r_i$ to be the minimum of the de Broglie
wavelength (the length scale over which wave dark matter is roughly
coherent or homogeneous \cite{Hui:2021tkt}), and 
and the radius of impact: 
\begin{equation}
\frac{r_i}{\rs}=\min\biggl\{\underbrace{10^3(\mu'\rs)^{-1}\biggl(\frac{v_{\rm typical}}{\SI{300}{\km/s}}\biggr)^{-1}}_{\text{de Broglie wavelength}},\underbrace{10^6\biggl(\frac{v_{\rm typical}}{\SI{300}{\km/s}}\biggr)^{-2}}_{\text{virial radius}}\biggr\}.
\label{eqn:r_i}
\end{equation}
The motivation for considering the de Broglie wavelength will be clear
in a moment.

Once the asymptotic dark matter density $\rho_i\approx T_{00}\approx
2\mu'{}^2|\Phi_{\rm acc} (r_i)|^2\approx \mu'{}^2|R_{\rm acc}
(r_i)|^2/(2\pi)$ is fixed,\footnote{For the last equality, we took the average over the angular variables.} equations (\ref{eqn:mass-evolution-single-mode-approx}) and (\ref{eqn:J-evolution-single-mode-approx}) supplied with (\ref{eqn:guanhaos-fit}) determine the dark matter accretion onto the black hole. Our goal is to study the subsequent evolution, and especially its interplay with superradiance. Note that equations (\ref{eqn:guanhaos-fit}) and (\ref{eqn:r_i}) are only needed to fix the normalization of the accreting flux, and thus the timescale of the threshold drift. A different normalization would not impact in any way the trajectory of the black hole in the Regge plane, nor the mass of the cloud as function of that of the black hole.

What values of $\ell'$ and $m'$ should we use for the accreting mode?
In general, several modes will be present at the same time, with a
distribution depending on the mass of the scalar and its velocity
dispersion. However, in two specific cases we can make a simplifying
assumption. 

One possibility is to assume spherical accretion with $\ell'=m'=0$. 
This can be motivated two different ways. 
For small values of $\mu'\rs$, the angular momentum barrier strongly
suppresses all modes with $\ell'\ge1$, see second and third line of
(\ref{eqn:guanhaos-fit}). It is thus natural to only consider the
purely radial infall corresponding to $\ell'=m'=0$.
Another motivation is the tendency of wave dark matter, especially in
the ultralight regime, to form solitons at the center of galaxy
halos \cite{Schive:2014hza}. The solitons provide a natural $\ell'=m'=0$ environment in
which the central supermassive black hole resides. 
A number of recent papers explore the interaction between a
supermassive
black hole and the soliton that hosts it
\cite{Hui:2019aqm,Chavanis:2019bnu,Davies:2019wgi,Padilla:2020sjy,Cardoso:2022nzc}.

A second possibility is to assume an accretion flow with
$\ell'=m'=1$. This is motivated by the observation that wave dark
matter generically has vortices, at the frequency of one vortex ring per de
Broglie volume \cite{Hui:2020hbq}. A vortex is a one-dimensional
structure, along which the dark matter density vanishes and around
which the dark matter velocity circulates, with winding number $m'$
generically being $\pm 1$. Let us thus consider a situation in which a
black hole happens to coincide with such a vortex, with $\ell'=m'=1$.
In general, there is no reason the black hole spin direction and the
vortex angular momentum align. We will for simplicity assume so,
noting that they would tend to align if the black hole is spun up by
the dark matter accretion. The question of whether a vortex, once it
intersects a black hole, would remain stuck to it, is an interesting
one, which we leave for future work.

It is helpful to have an idea of what the mass accretion rate might be
for these two possible scenarios. For spherical accretion:
\begin{eqnarray}
\label{Mdot0}
\dot M_{\rm acc} \sim 400 {\, \rm M_\odot \,/ \,yr.} 
\left( {\rho_i \over 10 {\,\rm M_\odot \,/\, pc^3}} \right)
\left( {M_{\rm BH} \over 10^9 {\,\rm M_\odot} }\right)^2
\left( {r_i / r_s \over 10^6} \right)^{3/2}
\end{eqnarray}
for $\mu' r_s \, \gsim \, 0.5$. The displayed value for density $\rho_i$ corresponds to that of
a soliton of mass $1.12\times 10^9 {\,\rm M_\odot}$ and
$\mu' = 10^{-22}$ eV. For $\ell' = m' = 1$ accretion:
\begin{eqnarray}
\label{Mdot1}
\dot M_{\rm acc} \sim 1.2 \times 10^{-2} {\, \rm M_\odot \,/ \,yr.} 
\left( {\rho_i \over 10 {\,\rm M_\odot \,/\, pc^3}} \right)
\left( M_{\rm BH} \over 10^9 {\,\rm M_\odot} \right)^2 
\left( {r_i/r_s \over 10^3} \right)^{3/2}\, 
\end{eqnarray}
in the particle regime $\mu' r_s \, \gsim \, 1$. There is suppression due to the angular momentum
barrier if the particle mass is low $\mu' r_s \, \lsim \, 1$ (for $\ell' =
1$), in which case the accretion rate becomes
\begin{eqnarray}
\label{Mdot2}
\dot M_{\rm acc} \sim 2.5 \times 10^{-5} {\, \rm M_\odot \,/ \,yr.} 
\left( {\rho_i \over 10 {\,\rm M_\odot \,/\, pc^3}} \right)
\left( M_{\rm BH} \over 10^9 {\,\rm M_\odot} \right)^2 
\left( {r_i/r_s \over 10^3} \right)^{3/2}
\left({\mu' r_s \over 0.5}\right)^9
\, .
\end{eqnarray}
Thus we see that the timescale for mass accretion, $M_{\rm BH}/\dot M_{\rm acc}$,
tends to be quite long (longer than a Hubble time) if
the dark matter accretion flow has non-vanishing angular momentum.

When (\ref{eqn:guanhaos-fit}) is substituted into either
(\ref{eqn:mass-evolution-single-mode-approx}) or
(\ref{eqn:J-evolution-single-mode-approx}), and the threshold
condition $\mu = m\Omega_+ = am/(r_s r_+)$ (i.e., we are studying
threshold drift at the superradiance threshold for $m$)
is imposed together with the conservation of the total mass, we get the following equations for the evolution of the mass of the black hole and of the cloud:
\begin{equation}
\frac{1-x^2}{(1+x^2)^2}\frac{\rd x}{\rd t}=K\frac{(m'-m\mu'/\mu)^2}{1+1/x^2},
\label{eqn:dx-dm}
\end{equation}
\begin{equation}
\frac{\rd(x+x_c)}{\rd t}=K\frac{(m\mu'/\mu-m')}{1+1/x^2}\frac{m\mu'}\mu,
\label{eqn:dxc-dm}
\end{equation}
where, as in Section \ref{sec:accretion+superradiance}, we defined
$x=\mu\rs/m$ and $x_c=(M_c/M_{\rm BH})x$. In these expressions, we set
$K\equiv4G|R_{\mathrm{acc}}(r_+)|^2(\mu/m)$, where
$R_{\mathrm{acc}}(r_+)$ is related to the asymptotic dark matter
density by (\ref{eqn:guanhaos-fit}).
To derive the above, we have set $\dot J_{\mathrm{acc}}/\dot
M_{\mathrm{acc}} = m'/\mu'$ in
Eqs.~(\ref{eqn:accretion-effective-x-evolution}) and
(\ref{eqn:accretion-effective-xc-evolution}), thanks to the
simplifying assumption that the wave dark matter accretion is due to a
single $m'$ mode. We have also used Eqs.~(\ref{eqn:mass-evolution-single-mode-approx}) and
(\ref{eqn:J-evolution-single-mode-approx}) which determine the mass
and angular momentum fluxes (with the unprimed $\mu$ and $m$ replaced
by $\mu'$ and $m'$). The relation $a^2 + r_+^2 = r_s r_+$ was useful
for relating $r_s/r_+$ to $x \equiv \mu r_s / m = a/r_+$ (assuming we are at
superradiance threshold for $m$), giving us $r_s/r_+ = 1 + x^2$.

From \eqref{eqn:dx-dm} we see that the mass of the black hole can only
increase (recall that $x < 1$; see Eq.~\ref{supercondition2}); on the
other hand, the sign of $\rd x_c/\rd t$ can be read off from
\eqref{eqn:over-under-superradiant}, which is
\begin{equation}
\frac{\rd x_c}{\rd t}=\biggl(\frac{1}{1-{\cal
    R}}\frac{1-x^2}{(1+x^2)^2}-1\biggr)\frac{\rd x}{\rd t},\qquad
{\cal R}\equiv\frac{m'\mu}{m\mu'},
\label{eqn:dxc/dt}
\end{equation}
which can be integrated to give a simple relation between the mass of the cloud and the mass of the black hole:
\begin{equation}
x_c+x-\frac{1}{1-{\cal R}}\frac{x}{x^2+1}=\text{constant}.
\label{eqn:xc(x)}
\end{equation}
This relation eliminates the time dependence and gives the mass of the
cloud as function of the mass of the black hole, $x_c(x)$. 
It tells us that the total mass of the black hole + cloud system is
determined by a constant (fixed by initial conditions for $x$ and
$x_c$ on the superradiance threshold) plus $x/[(1 - {\cal R}) (x^2 + 1)]$.

{Three main different cases can be distinguished.}

\begin{description}
\item[Case 1.] If ${\cal R} \equiv m'\mu/(m\mu') \le0$, then the parenthesis on the right-hand side of (\ref{eqn:dxc/dt}) is always negative for $0<x<1$.
This is under-superradiance: the mass of the cloud
\textit{decreases}, $\rd x_c/\rd t\le0$, while the mass of the black hole
increases, $\rd x/\rd t \ge 0$ (recall that on the threshold, 
the black hole can only move to the
right in the Regge plane, by the second law).
Moreover, the black hole's mass will increase faster than
the ambient accretion rate, i.e.~$\rd x/\rd
t\ge(\mu/m)[\rd r_s/\rd t]_{\mathrm{acc}}$, because it receives
mass from both the ambient environment and the cloud that
was built up by superradiance.
The black hole will thus
move along the superradiance threshold faster than naively expected
based on ambient accretion alone,
but only before the cloud is depleted.  An example of this case is
shown in Figure~\ref{fig:dm-accretion-R=0}, for the case of
spherically-symmetric accretion $\ell'=m'=0$. After the cloud's
depletion, the black hole will leave the threshold and move under the
effect of accretion alone. How much faster can the black hole increase its mass? From (\ref{eqn:dxc/dt}), we find that the speed is increased by a factor
\begin{equation}
\frac{\dot x}{\dot x_{\rm acc}}=(1-{\cal R})\frac{(1+x^2)^2}{1-x^2},
\end{equation}
which can in principle become very large near the edge of the
superradiant threshold, around $x\lesssim1$, where the black hole
approaches extremality. ($\dot x_{\rm acc} \equiv (\mu/m) \rd r_s/\rd t
|_{\rm acc} = \dot x + \dot x_c$.) Remember however that this would also mean
that the cloud depletes very fast, and thus that the threshold drift
will end very soon.

\begin{figure}
	\centering
	\includegraphics[width=0.85\textwidth]{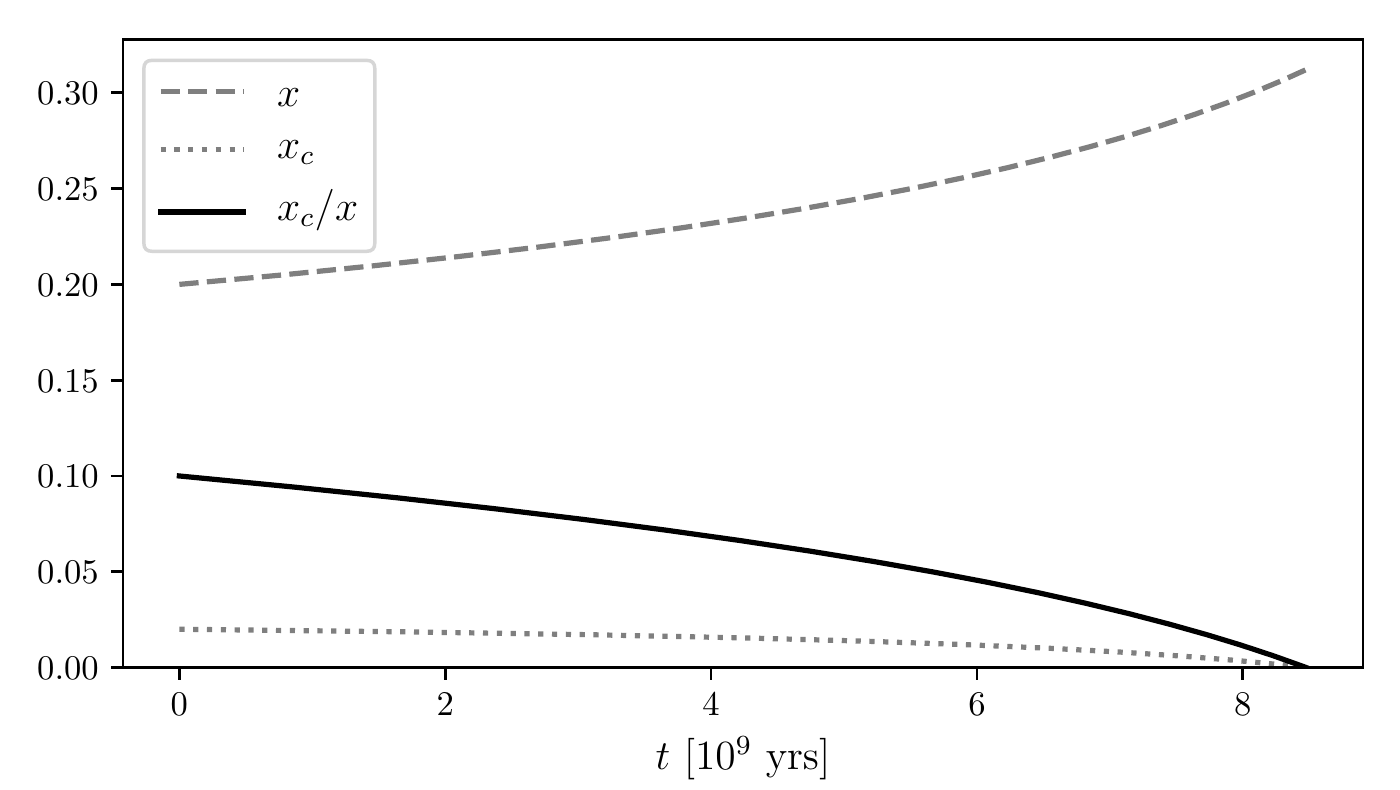}
	\caption{Evolution of the mass of the black hole and of the
          cloud, solving (\ref{eqn:dx-dm}) and (\ref{eqn:dxc-dm}) with
          $m'=0$ (spherically symmetric accretion) and $m=1$. The
          masses are expressed in dimensionless units as 
          $x=\mu\rs/m$ and $x_c = x
          M_{c}/M_{\rm BH}$. Formulae (\ref{eqn:guanhaos-fit}) and
          (\ref{eqn:r_i}) have been used to determine the amplitude of
          the field close to the black hole from $\rho_i$ and $v_{\rm
            typical}$, which we fixed to be $\SI{e+3}{GeV/cm^3} = 27
          {\,\rm M_\odot/pc^3}$ and
          $\SI{300}{km/s}$ respectively. The mass of the scalar
          field(s) has been fixed to $\mu=\mu'=\SI{e-20}{eV}$, while
          the initial value of $x$ is $0.2$ (corresponding to
          $1.34\times10^9\,M_\odot$) and that of $x_c$ is $0.02$.
        This is an example of under-superradiance (Case 1), where a cloud that
        was built up from superradiance ($m=1$) gives mass back to the black
        hole, while ambient accretion ($m'=0$) also adds mass to the black hole.}
	\label{fig:dm-accretion-R=0}
\end{figure}

\item[Case 2.] If $0< {\cal R} \equiv m' 
\mu / (m \mu') <1$, then we have over-superradiance for $x<x_\star$ and under-superradiance for $x>x_\star$, where $x_\star$ is the zero of (\ref{eqn:dxc/dt}):
\begin{equation}
\frac{1-x_\star^2}{(1+x_\star^2)^2}=1-{\cal R}\implies
x_\star=\sqrt{\frac{-3+2{\cal R}+\sqrt{9-8{\cal R}}}{2(1-{\cal R})}}.
\label{eqn:x_star-result}
\end{equation} 
The point $x=x_\star$ (depicted with a star in 
Figure \ref{fig:dm-accretion-super}; $x_\star \sim 0.5$ for 
${\cal R} = 1/2$)
corresponds to a local maximum for the mass of
the cloud during threshold drift. 
This case is perhaps the most
interesting one, because over-superradiance provides a mechanism to 
boost the cloud's mass. 

\begin{figure}
	\centering
	\includegraphics[width=0.75\textwidth]{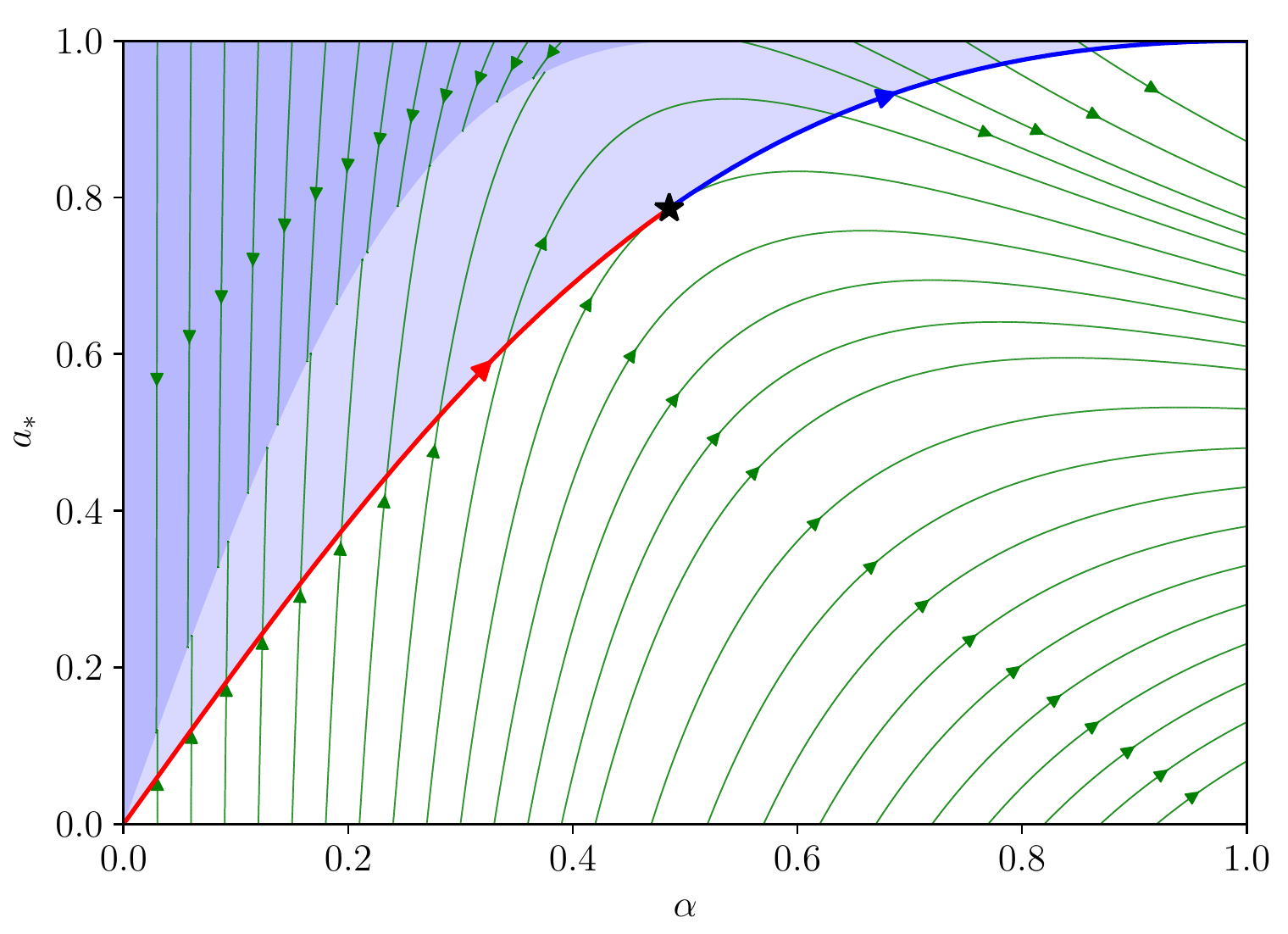}
	\caption{Regions of superradiance instability for $m=1$ (blue)
          and $m=2$ (light blue) in the Regge plane, together with the
          streamlines of dark matter accretion (from
          \eqref{eqn:mass-evolution-single-mode} and
          \eqref{eqn:J-evolution-single-mode}, in green), for $m'=1$
          and $\mu'=\mu$. When an $m=2$ cloud is present, the black
          hole moves along the corresponding threshold: the mass of
          the cloud increases along the red line (over-superradiance),
          reaches a maximum at the star (see equation
          (\ref{eqn:x_star-result})) and then decreases along the blue
          line (under-superradiance). Throughout both processes, the
          black hole's mass increases. The threshold drift depicted
          here is an illustration of Case 2, combining $m=2$
          superradiance cloud with $m'=1$ ambient accretion.}
	\label{fig:dm-accretion-super}
\end{figure}

\begin{figure}
	\centering
	\includegraphics[width=0.9\textwidth]{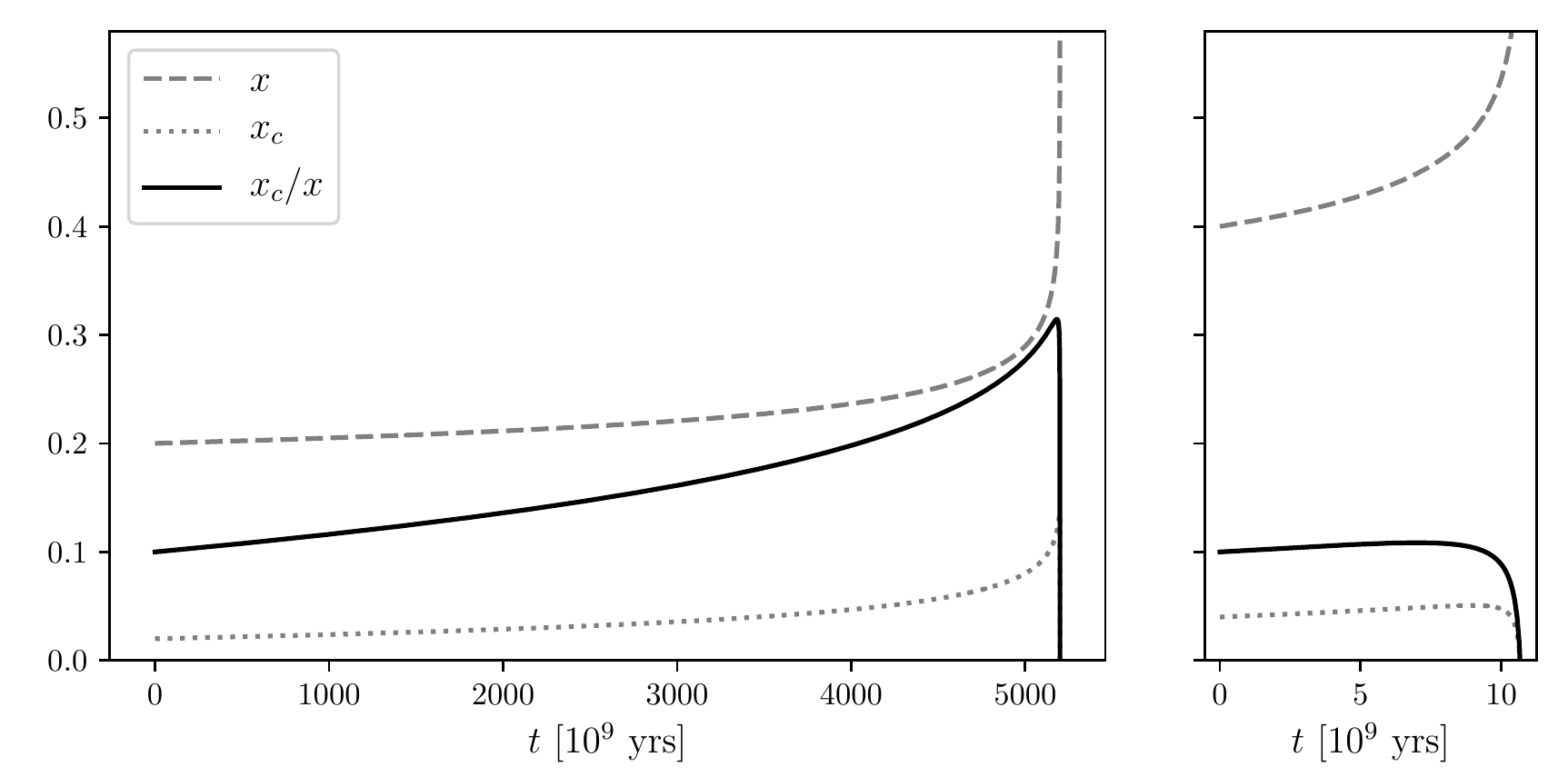}
	\caption{An illustration for Case 2: evolution of the mass of the black hole and of the
          cloud, solving (\ref{eqn:dx-dm}) and (\ref{eqn:dxc-dm}) with
          $m'=1$ (ambient vortex accretion) and $m=2$ superradiance
          cloud. In the left panel, the parameters are the same as for
          Figure~\ref{fig:dm-accretion-R=0}; in the right panel, the
          initial value of $x$ is taken to be $0.4$, to illustrate the
          dependence on initial condition. In the left panel, the
          evolution of the black hole requires a time much longer than
          the age of the Universe, as a consequence of (1) the
          expected modest dark matter density in a typical environment,
          and (2) the angular momentum suppression in the
          ``Intermediate'' regime in (\ref{eqn:guanhaos-fit}).
         This example is thus not of practical interest, but serves
         the purpose of demonstrating the boost of cloud-to-black-hole mass
       ratio by over-superradiance. We will see an example of more
       practical interest with baryonic accretion.}
	\label{fig:dm-accretion-R=0.5}
\end{figure}

Note that the cloud-to-black hole mass ratio,
\begin{equation}
\frac{x_c}x=\frac1x\biggl(x_c(0)+x(0)-\frac1{1-{\cal
    R}}\frac{x(0)}{1+x(0)^2}\biggr)-1+\frac1{1-{\cal R}}\frac1{1+x^2} \, ,
\label{eqn:x_cx}
\end{equation}
reaches maximum at a certain $x<x_\star$. Depending on the initial
masses $x(0)$ and $x_c(0)$, as well as on the value of ${\cal R}$,
this cloud-to-black hole mass ratio can reach higher values than the
10\% mentioned in Section~\ref{sec:single-evolution}. We show an
example in Figure~\ref{fig:dm-accretion-R=0.5}, where we use the same
parameters as in Figure~\ref{fig:dm-accretion-R=0}, but change the
values of $m'$ and $m$ to 1 and 2 respectively. First of all, we see
that, even if the asymptotic dark matter density is the same in both
cases, the evolution is now much slower, due to the suppression by
angular momentum as indicated by (\ref{eqn:guanhaos-fit}). Second, we observe
that the cloud's mass is indeed boosted, reaching values as high as
$x_c= 0.31 \, x$ at the peak, beyond the limit of about $0.1$ derived
in (\ref{eqn:limit-analytical}), though this requires a
long timescale ($\sim 5 \times 10^{12}$ yrs in Figure
\ref{fig:dm-accretion-R=0.5}), because typical dark matter density in
the environment gives only a somewhat low accretion rate.
Higher values of $x_c$ are possible, but require starting the
evolution at lower values of $x(0)$, so that the black hole drifts
along the superradiant threshold for a longer time, giving the cloud
more time to grow. In fact, from (\ref{eqn:x_cx}), the highest
possible cloud-to-black-hole mass ratio is $\mathcal R/(1-\mathcal R)$
and is formally attained in the $x(0) \rightarrow 0$ limit.
For the parameters chosen for Figure~\ref{fig:dm-accretion-super}, 
${\cal R} = 1/2$ which means the cloud-to-black-hole mass ratio
could in principle reach unity.
Because (\ref{eqn:guanhaos-fit}) is suppressed for
small values of the black hole mass, however, this would require
waiting for an exponentially longer time before reaching the peak. In the
right panel of Figure~\ref{fig:dm-accretion-R=0.5}, we show that the
evolution is faster for larger $x(0)$, but also that the cloud's mass
cannot grow as large. 

The upshot is that the potentially large cloud
mass that could be attained by threshold drift is a bit academic, if
the source of ambient accretion is dark matter, due to its
modest density in typical environments, resulting in a long
timescale for cloud build-up. 
However, we will see below the case of ambient accretion from a
baryonic disk, which gives a much shorter timescale.

\item[Case 3.] If ${\cal R} \equiv m'\mu/(m \mu') >1$, then from
  equation \eqref{eqn:dxc-dm} we see that the 
total (black hole + cloud) accretion rate is negative. This means that
the ``ambient accretion'' is actually not accreting at all, but is itself in
the regime of superradiance, extracting mass and angular momentum
from the combined black hole + cloud system, i.e.~it is more accurate to call
it ambient extraction. This is not bound
superradiance, as in
the case of the superradiance cloud, but unbound
superradiance (i.e., entries 1 and 3 respectively in Table~\ref{tab1}).
However, recall from \eqref{eqn:dx-dm} that the
black hole cannot lose mass while moving along the threshold. This is not
a contradiction: it means the presence of an external extraction of
mass (and angular momentum) induces the already existing cloud to lose
to the black hole more mass (and angular momentum) than what is
extracted. In other words, what we have is under-superradiance:
the system as a whole (black hole + cloud) loses mass; the cloud loses
mass faster than the whole system; the black hole gains
mass.

This case with ${\cal R} \equiv m'\mu/(m \mu') >1$ can be achieved for
instance by $m' = m = 1$ and $\mu > \mu'$.
If $\mu'=\mu$, this threshold drift by ambient extraction requires
$m'\ge2$ since $m$ has to be at least unity. Winding higher than unity
is not generically expected for a vortex formed out of chance
destructive interference in wave dark matter \cite{Hui:2020hbq}.
However, in Section \ref{sec:transition} we will see how this case
can be realized fairly naturally, by replacing the unbound
superradiance (from the ambient environment) with bound superradiance
(from another level $m' \ne m$ in the superradiance cloud).
\end{description}

\subsection{Level transition}
\label{sec:transition}

\begin{figure}
	\centering
	\includegraphics[width=0.85\textwidth]{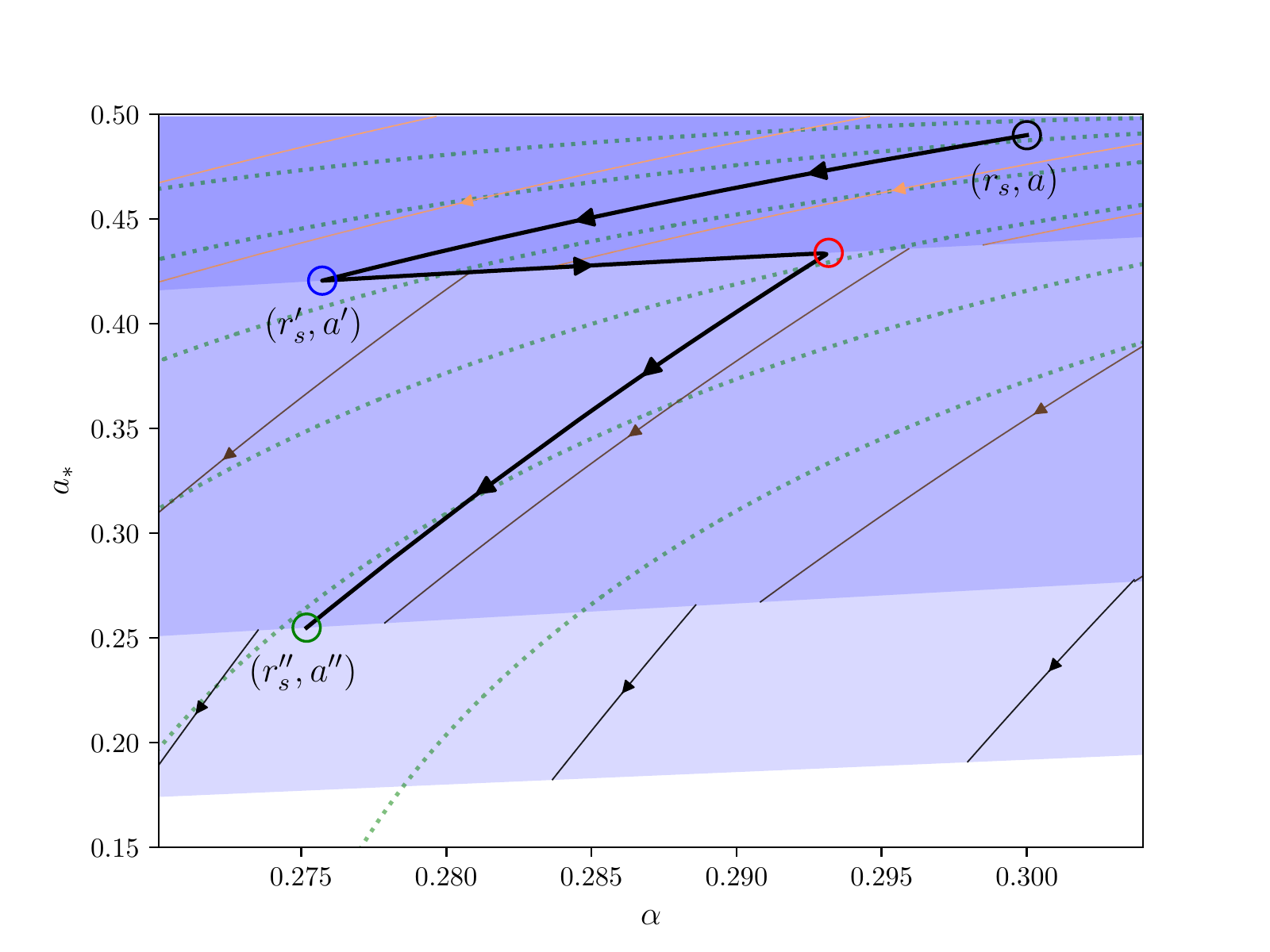}
	\caption{Trajectory in the Regge plane followed by an isolated
          black hole under the effect of superradiance from two modes
          (thick black lines), computed from equations (36) of
          \cite{Ficarra:2018rfu}. The two modes are $m=1$ and $m'=2$.
          The trajectory has a zig-zag shape, whose
          endpoints' positions (blue, red and green circles) we
          predict analytically. The portion from blue circle to red
          circle is level transition, which can be understood as
          threshold drift along the $m=1$ superradiance threshold.
          The green circle is located on the $m'=2$ threshold.
          The green dotted lines are lines of equal horizon area.
          The thin lines with arrows indicate generic trajectories
          for modes with $m'=1,2,3$.}
	\label{fig:transition}
\end{figure}

The evolution of mass and spin of an isolated black hole, due to the superradiance instability, is
initially dominated by the fastest-growing mode. Let us use $m$ to
denote the angular momentum of this mode.
This phase of the evolution, described in Section \ref{sec:single-evolution}, brings the black hole from its initial position in the Regge plane (black circle in Figure~\ref{fig:transition}) to the threshold of the instability region of the grown mode (blue circle) along a trajectory determined by
\begin{equation}
\frac{\rd(a\rs)/\rd t}{\rd\rs/\rd t}=\frac{m}\mu.
\end{equation}
The final parameters ($\rs',a')$ are linked to the initial ones ($\rs,a)$ by equation (\ref{eqn:analytic-formulae}).
If other modes (and the ambient environment) are neglected, the system
is now in stable equilibrium, with the black hole sitting on the level
$m$ superradiance threshold.

What happens when other modes (of the bound cloud) are taken into account? The exponential dependence of the instability rate on the angular momentum number, $\Im(\omega)\sim\alpha^{4\ell+5}$, ensures a large separation of timescales between the growth of the first and of the next levels. The next-fastest growing mode will thus `silently' grow without noticeably affecting the black hole parameters for a long time, until its mass eventually becomes comparable to that of the, previously grown, fastest mode.

Given this large separation of timescales, we can describe the
evolution of the system with the approach developed in
Section~\ref{sec:accretion+superradiance}, similar to ``Case 3'' of
Section \ref{sec:dm-accretion}. The main difference from Case 3 there
is this: the mass and spin extracted by the
next-fastest growing mode (let us denote its angular momentum by $m'$),
do not escape to infinity, but rather go into the $m'$ level of the bound superradiance cloud.
The black hole undergoes threshold drift, from the blue to the red
circle as in Figure~\ref{fig:transition}), according to equation
(\ref{eqn:xc(x)}) with $\mu'=\mu$. 
The drift is of the under-superradiance type, just as in
Case 3 before, such that the level $m$ part of the cloud
loses mass to the black hole, while the level $m'$ part gains
mass from the black hole, and the black hole as a whole
gains mass. The threshold drift stops when level $m$
is completely depleted, indicated by the position of the red circle
in Figure~\ref{fig:transition}. The threshold drift from blue circle to red
is what we call {\it level transition}.\footnote{{Note that this is different from the ``atomic level transition'' that accompanies the emission of gravitational waves, see e.g.~\cite{Arvanitaki:2014wva}.}} The superradiance cloud
switches from being dominated by level $m$ to being dominated by level
$m'$.

Once level $m$ is emptied out,  level $m'$ continues its
superradiance growth on its own. Without level $m$, there is no
longer the glue to keep the black hole stuck at the level $m$
threshold. Thus, the black hole moves from red to green circle. 
This is just the standard single mode Regge trajectory. The green
circle is where the trajectory hits the level $m'$ threshold.
The trajectory is determined by
\begin{equation}
\frac{\rd(a\rs)/\rd t}{\rd\rs/\rd t}=\frac{m'}\mu,
\label{eqn:trajectory-m'}
\end{equation}
The whole evolution from the initial black circle to the final green
circle is thus a zig-zag in the Regge plane.

The locations of all the colored circles in the Regge plane can be
written down analytically given the initial (black) circle.
The blue circle is given by (\ref{eqn:analytic-formulae}). 
The green circle is given by the same expression with 
$m \rightarrow m'$, $a' \rightarrow a''$, $r_s' \rightarrow r_s''$:
\begin{equation}
\frac{\mu\rs''}{m'}=\frac{1-\sqrt{1-(2(\mu\rs/m')(1-\mu a/m'))^2}}{2(\mu\rs/m')(1-\mu a/m')},\qquad \frac{a''}{\rs''}=\frac{\mu\rs}{m'}\Bigl(1-\frac{\mu a}{m'}\Bigr).
\label{eqn:rs''a''}
\end{equation}
Notice how, because total mass and angular momentum are conserved,
the green circle is no different from where the black hole would end
if there were only level $m'$ superradiance all along.
The intersection of such a Regge trajectory with the level $m$
threshold gives the location of the red circle.

\begin{figure}
	\centering
	\includegraphics[width=0.9\textwidth]{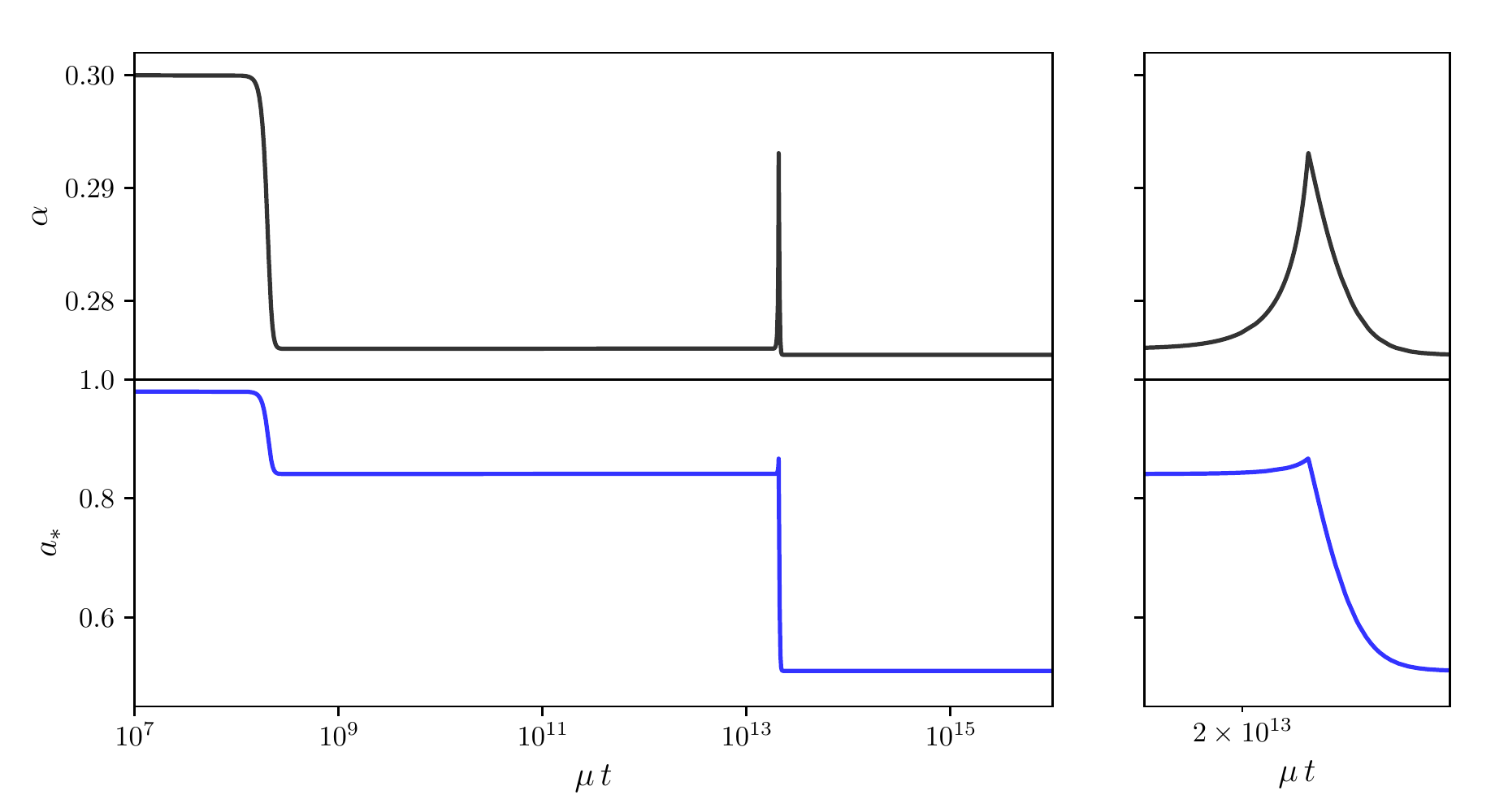}
	\caption{\textit{Left panel:} mass and spin evolution of the black hole during the trajectory depicted in Figure \ref{fig:transition}. \textit{Right panel:} zoom around the level transition.}
	\label{fig:transition-in-time}
\end{figure}

All these conclusions can be verified by computing the time evolution
of the black hole parameters with the two-mode model developed in
\cite{Ficarra:2018rfu}. The black line in Figure~\ref{fig:transition}
is the result of a numerical integration of equations (36) of
\cite{Ficarra:2018rfu} for the case with $\ell=m=1$ and $\ell'=m'=2$,
with small initial seeds for both modes. Its zig-zag shape is evident,
as well as its perfect match with the position of the circles, which
are computed with our analytical description of the evolution. The
results as a function of time are reported in Figure~\ref{fig:transition-in-time}.

\subsection{Baryonic accretion}
\label{sec:baryonic}

A natural kind of ambient accretion around a black hole is due to a
baryonic accretion disk. Its interplay with the superradiance was
considered in \cite{Brito:2014wla}. The results of
\cite{Brito:2014wla} showed that, for a significant portion of its
evolution, the black hole evolved along the superradiance threshold,
in the sense we described in
Section~\ref{sec:accretion+superradiance}. In this section, we show
how their results can be understood in a simple way by considering (\ref{eqn:accretion-effective-x-evolution}) and (\ref{eqn:over-under-superradiant}).

The accretion rate considered in \cite{Brito:2014wla} is a fraction $f_{\rm Edd}$ of the Eddington rate\footnote{Here, $\sigma_{\rm T}$ is the Thomson cross section and $m_p$ is the proton's mass.} \cite{Barausse:2014tra,Lynden-Bell:1969gsv,Soltan:1982vf},
\begin{equation}
\dot M_{\rm acc}=\frac{f_{\rm Edd}}{\tau_{\rm Sal}}M_{\rm BH},\qquad \tau_{\rm Sal}=\frac{2\Mpl^2\sigma_{\rm T}}{m_p}=\SI{4.5e+7}{yrs},
\label{eqn:dotMacc_baryonic}
\end{equation}
with the angular-momentum-to-mass accretion ratio given by that of particles on the ISCO \cite{Bardeen:1970zz},
\begin{equation}
\frac{\dot J_{\rm acc}}{\dot M_{\rm acc}}=\frac{\rs}{3\sqrt3}\frac{1+2\sqrt{3r_{\rm ISCO}/GM-2}}{\sqrt{1-2GM/(3r_{\rm ISCO})}},
\label{eqn:dotJacc_baryonic}
\end{equation}
where $r_{\rm ISCO}$ is the radius of the innermost stable circular
orbit. It is straightforward to implement these formulae in the
effective equations derived in
Section~\ref{sec:accretion+superradiance}.

It should be noted that the above accretion model
has certain limitations. For instance, $f_{\rm Edd}$ is likely
a function of time. Also, we expect modifications to
$\dot J_{\rm acc} / \dot M_{\rm acc}$ as the black hole gets
spun up close to extremality (\cite{Thorne:1974ve} gave an upper bound of
$a_*\sim 0.998$).

\begin{figure}
	\centering
	\includegraphics[width=0.8\textwidth]{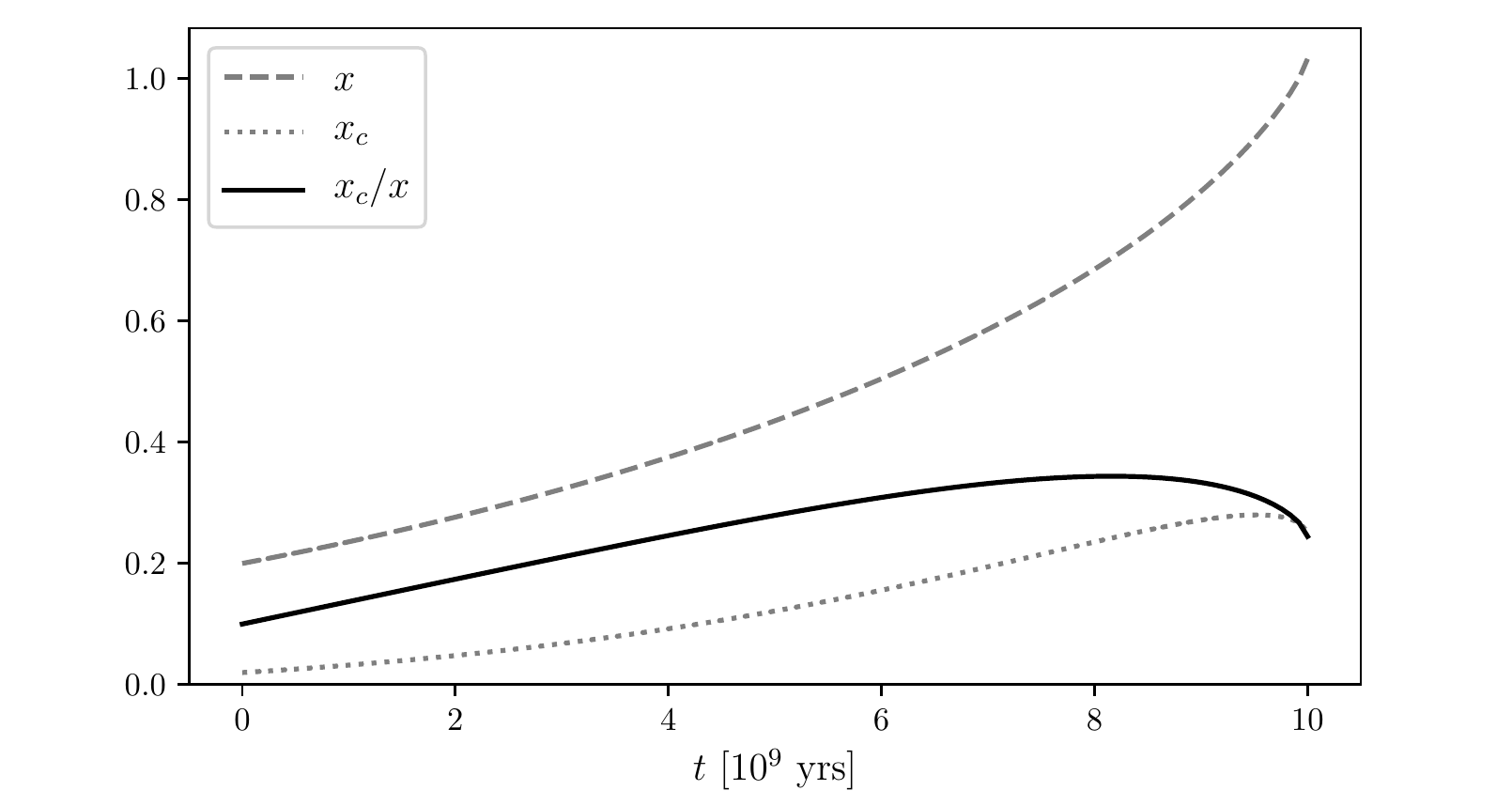}
	\caption{Evolution of a black hole + superradiance cloud
          system, in the presence of accretion from a disk: solution of equations
          (\ref{eqn:accretion-effective-x-evolution}) and
          (\ref{eqn:accretion-effective-xc-evolution}) with accretion
          given by (\ref{eqn:dotMacc_baryonic}) and
          (\ref{eqn:dotJacc_baryonic}) with $f_{\rm Edd}=0.01$. The
          associated threshold drift proceeds
          until the black hole hits extremality,
          corresponding to $x=1$; however, as explained in the text,
          equations (\ref{eqn:accretion-effective-x-evolution}) and
          (\ref{eqn:accretion-effective-xc-evolution}) break down
          close to that point. As shown by the dotted line in the
          plot, the cloud's mass increases during the threshold drift,
          i.e.~this is over-superradiance.
          See footnote \ref{cloudcorrections} on the slight dip in
          cloud mass towards the end.
}
	\label{fig:bar-accretion}
\end{figure}

We solve (\ref{eqn:accretion-effective-x-evolution}) and
(\ref{eqn:over-under-superradiant}), with $f_{\rm Edd} = 0.01$,
an initial black hole mass of $x=0.2$ and cloud mass of $x_c = 0.1 \, x$.
(Recall $x \equiv \mu r_s/m$, $x_c \equiv M_c x / M_{\rm BH}$; the
precise value for $\mu/m$ does not matter once one expresses
everything in terms of $x$ and $x_c$.) The results are shown in Figure~\ref{fig:bar-accretion}. The threshold drift associated with baryonic accretion is like that depicted earlier in Figure~\ref{fig:dm-accretion-super}, except that the entire drift is in the over-superradiance regime: the cloud's mass in Figure~\ref{fig:bar-accretion} increases with
time.
In other words, there is no under-superradiance portion, and over-superradiance continues all the way until the black hole is close to extremality.\footnote{\label{cloudcorrections} Close to the end, at high masses and spins,
  the cloud's mass can be seen decreasing slightly. 
  This is an artifact of the $\omega \approx \mu$ approximation
  used in deriving (\ref{eqn:accretion-effective-x-evolution}) and
  (\ref{eqn:over-under-superradiant}). We have checked that including
  higher-order corrections in $\alpha$ goes in the direction of
  restoring over-superradiance at the very end of the cloud's
  evolution. One should also keep in mind, as remarked above,
  additional effects could prevent the black hole from reaching extremality.
}
It is easy to see that the smaller the initial mass of the black hole, the larger the $x_c/x$ ratio can grow. By using this observation, we are able to find the largest achievable $x_c/x$ ratio, which happens to be roughly 36\%. This extends the results of \cite{Brito:2014wla}, where an example with $x_c/c$ larger than 30\% was shown.

It is interesting to note that, if the accretion disk produces a strong magnetic field, the Blandford–Znajek (BZ) process \cite{Blandford:1977ds} can contribute to the energy extraction from the black hole, alongside superradiance. This effect can in principle simply be added as an extra source term in (\ref{eqn:accretion-effective-x-evolution}) and
(\ref{eqn:over-under-superradiant}), contributing with sign opposite to that of baryonic accretion. The impact of the BZ process on the mass-spin evolution of the black hole, however, is expected to be negligible compared to the accretion from the disk. As a rough comparison, from well-known estimates of its power output $\dot M_{\rm BZ}$ (see, e.g., chapter IV of \cite{Thorne:1986iy}), we find
\begin{equation}
\frac{\dot M_{\rm BZ}}{\dot M_{\rm acc}}\sim\frac{7\times10^{-4}}{f_{\rm Edd}}\biggl(\frac{M_{\rm BH}}{10^9M_\odot}\biggr)\biggl(\frac{B_\perp}{\SI{e+4}{G}}\biggr)^2,
\label{eqn:bz}
\end{equation}
where $B_\perp$ is the normal magnetic field at the horizon. The estimate (\ref{eqn:bz}) assumes extremal black hole spin, for which $\dot M_{\rm BZ}$ is maximized.

\section{Discussion}
\label{sec:discuss}

In summary, we have explored how superradiance (which extracts mass
and angular momentum from a black hole) could work in tandem with 
accretion (which donate both to the black hole).
Superradiance, because of its ability to build up a substantial cloud
around the black hole, is often more efficient than accretion from the
ambient environment, see discussion at the end of Section
\ref{sec:single-evolution}, around Eqs.~({\ref{Mdot0})-(\ref{Mdot2})
and Eq.~(\ref{eqn:dotMacc_baryonic}).

This means the black hole will generically evolve
towards the superradiance threshold in the Regge (black hole spin
versus mass) plane. Once sufficiently close to the threshold, the
superradiance rate is reduced to an extent that accretion can
compete. The subsequent evolution of the black hole spin and mass,
a climb along the threshold we call threshold drift, is the focus of
this paper. We provide simple evolution equations describing the
climb: Eqs.~(\ref{eqn:dx-dm}) and (\ref{eqn:dxc/dt}).
We give an analytic relation between the black hole mass and superradiance
cloud mass (\ref{eqn:xc(x)}), and a formula for the end-point of
over-superradiance (\ref{eqn:x_star-result}).

Of the possible scenarios, perhaps the most interesting ones are cases where
$\mu/m$ (mass-to-angular-momentum ratio for the superradiance scalar)
is less than $\dot M_{\rm acc}/\dot J_{\rm acc}$ (mass-to-angular-momentum accretion rate
ratio), assuming both have the same sign as the black hole spin. 
The black hole {\it gains} mass and angular momentum during 
the threshold drift, even as the superradiance cloud does {\it the
  same}. Effectively, the ambient accretion serves to feed the
superradiance cloud {\it via} the black hole. We refer to this process as
over-superradiance. This way, the
superradiance cloud can acquire a mass that exceeds the standard
maximum of $10\% M_{\rm BH}$ from superradiance
alone without accretion. We have considered two separate examples of
ambient accretion: one is accretion of the surrounding dark matter
from a wave dark matter vortex (with $\dot J_{\rm acc}/\dot M_{\rm
  acc} = m'/\mu'$ and $m'=1$; Case 2 of Section \ref{sec:dm-accretion}); the other is the accretion of baryons
from a disk (Section \ref{sec:baryonic}). The latter is more efficient, and the highest
superradiance cloud mass we find is about $35 \%$  that of the black
hole, consistent with the results of \cite{Brito:2014wla}.
Dark matter accretion can in principle achieve an even higher
cloud mass (comparable to that of the black hole), with the caveat that the accretion rate is slow and the cloud
build-up takes longer than a Hubble time, see (\ref{Mdot1}) and
(\ref{Mdot2}). The long timescale for dark matter accretion is
due both to the moderate dark matter density in typical environments,
and to the suppression of accretion by the angular momentum barrier
(\ref{eqn:guanhaos-fit}). One could get around the angular
momentum suppression by increasing the dark matter particle mass
$\mu'$ (while keeping the superradiance scalar mass $\mu$ fixed).
But it can be shown the resulting cloud-to-black-hole mass ratio
is diminished (\ref{eqn:x_cx}), due to the smaller ${\cal R} \equiv m'\mu/(m\mu')$.

Dark matter accretion can proceed at a much faster rate for spherical
accretion, which is the example depicted in Figure
\ref{fig:dm-accretion-R=0} (Case 1 in Section \ref{sec:dm-accretion}).
The process illustrated there is under-superradiance, where the cloud
mass shrinks while both the cloud and the ambient accretion feeds the
black hole. Perhaps most interesting is the fact that the black hole
spins up during the threshold drift (it has to, by the second law),
despite the fact that the ambient accretion gives mass but not angular
momentum to the black hole. The spin-up of the black hole is entirely due to the angular
momentum from the diminishing cloud. 

The possibility of a substantial superradiance cloud raises a number
of interesting questions. (1)~The cloud's own gravity
cannot be ignored, that is to say, the geometry is no longer
completely dominated by the black hole. How will this affect the
dynamics and evolution of the cloud? It is known that a
self-gravitating, rotating boson cloud (without a black hole) is
unstable on short timescales
\cite{Sanchis-Gual:2019ljs,Dmitriev:2021utv}. 
How would accounting for both the gravity of the black hole and 
that of the cloud modify the story? As one dials up the
cloud-to-black-hole mass ratio, when will the instability
observed by \cite{Sanchis-Gual:2019ljs,Dmitriev:2021utv} become relevant? (See also \cite{Cardoso:2022nzc} for a recent numerical  solution of the accretion process of a boson star by a black hole.)
(2)~An increased mass of the cloud
  will, in general, enhance nonlinear effects such as
  self-interaction and the gravitational backreaction on the geometry. At a minimum, such nonlinear effects would change the profile of
the cloud and, possibly, the associated flux through the horizon. The relative importance
of the cloud's self-gravity versus the black hole's gravity
is obviously determined by $M_{c}/M_{\rm BH}$. For
self-interaction, the relevant self-interaction to gravity ratio 
is $\lambda \Phi^4 / (\mu^2 \Phi^2 r_s/r)$
where $r \sim 1/(r_s \mu^2)$ is the cloud size, and $\lambda$ is the
self-coupling strength (for an axion, $\lambda
\sim \mu^2/F^2$ where $F$ is the axion decay constant).
This ratio is roughly $\alpha^2 (M_{c}/M_{\rm BH}) (M_{\rm
  Pl}^2/F^2)$. Moreover, self-interaction is able to shutdown the
growth of subdominant superradiant modes via level mixing, as
explained in \cite{Arvanitaki:2010sy}, and also trigger scalar
emission \cite{Baryakhtar:2020gao}.
It would be useful to explore these nonlinear effects further \cite{Gruzinov:2016hcq}, in
light of the possibility of a substantial cloud mass, and thus cloud density.
(3)~In a binary setting, if one (or
both) of the binary components has a substantial cloud, the inspiral
dynamics can be heavily affected. 
For example, in cases of extreme
mass ratios, a small compact object can move through the cloud of the
big black hole. 
A more massive cloud will lead to enhanced dynamical friction,
accretion and 
orbital resonances \cite{Hui:2016ltb,Baumann:2018vus,Zhang:2019eid,Baumann:2021fkf,Baumann:2022pkl, Traykova:2021dua,Buehler:2022tmr,Boudon:2022dxi,Vicente:2022ivh}.
(4)~The threshold drift phenomenon implies that black
holes can experience interesting evolution along superradiance
thresholds. Under what circumstances is such an evolution observable in
real time, such as in Event Horizon Telescope data? 

These questions deserve further investigation. We hope to do so 
in the near future.

\section*{Acknowledgements}

We thank Dan Kabat and Ted Jacobson for useful discussions.
LH and GS are supported by the DOE DE-SC0011941 and a 
Simons Fellowship in Theoretical Physics. GS is supported in part by the U.S. Department of Energy Grant No. DE-SC0009919.
AL is supported in part by the Croucher Foundation and DOE grant de-sc/0007870.
ET is partly supported by the Italian MIUR under contract 2017FMJFMW (PRIN2017).

\appendix

\section{Scalar hair around Kerr black holes}
\label{app1}

In this appendix we collect relevant facts about scalar hair solutions around a Kerr black hole. We start by presenting the full exact solution to the Klein-Gordon equation, followed by analytic estimates for the particle and wave regimes. Finally we present numerical results to support the estimates in \eqref{eqn:guanhaos-fit}.

	\subsection{Klein-Gordon equation in a Kerr background} \label{KG kerr}
	
	The exact solutions to the Klein-Gordon equation in a Kerr-Newman background are constructed in \cite{Vieira:2014waa}. Here we restrict ourselves to the Kerr case ($Q=0$), with metric \eqref{kerr}. The Klein-Gordon equation for a (complex) scalar field $\Phi$ with mass $\mu$ in Boyer-Lindquist coordinates reads
	\begin{align}\label{KG Kerr}
		0=& \, \bigg\{ \frac{1}{\Delta} \Big[ (r^2+a^2)^2-\Delta a^2 \sin^2 \theta\Big]\frac{\partial^2}{\partial t^2}-\frac{\partial}{\partial r} \bigg(\Delta \frac{\partial}{\partial r}\bigg)-\frac{1}{\sin\theta}\frac{\partial}{\partial\theta}\bigg( \sin\theta\frac{\partial}{\partial\theta}\bigg)\nn\\
		&-\frac{1}{\Delta\sin^2\theta}(\Delta-a^2\sin^2\theta)\frac{\partial^2}{\partial\phi^2}+\frac{2a}{\Delta}\Big[(r^2+a^2)-\Delta\Big]\frac{\partial^2}{\partial t\,\partial\phi}+\mu^2 \varrho^2 \bigg\}\Phi\,,
	\end{align}
where $\varrho^2 \equiv r^2 + a^2 {\,\rm cos\,}^2\theta$ and $\Delta \equiv r^2 - rr_s + a^2 =  (r-r_+)(r-r_-)$ with $r_\pm \equiv
r_s/2 \pm \sqrt{(r_s/2)^2 - a^2}$.
	To solve \eqref{KG Kerr}, we make the ansatz 
	\begin{align}
		\Phi= e^{-i\omega t} e^{im\phi} S(\theta) R(r)\,.
	\end{align}
Substituting this into \eqref{KG Kerr} leads to
	\begin{align}
		0=& \, \frac{1}{\Delta} \Big[ (r^2+a^2)^2-\Delta a^2 \sin^2 \theta\Big](-\omega^2)-\frac{1}{R}\frac{\rd}{\rd r} \bigg(\Delta \frac{\rd R}{\rd r}\bigg)-\frac{1}{S}\frac{1}{\sin\theta}\frac{\rd}{\rd\theta}\bigg( \sin\theta\frac{\rd S}{\rd \theta}\bigg)\nn\\
		&-\frac{1}{\Delta\sin^2\theta}(\Delta-a^2\sin^2\theta)(-m^2)+\frac{2a}{\Delta}\Big[(r^2+a^2)-\Delta\Big](-i\omega)(im)+\mu^2 \varrho^2 .
	\end{align}
	We isolate the $r$- and $\theta$-dependent terms and pick the separation constant $\lambda$ such that the angular and radial equations are \cite{Teukolsky1973PerturbationsOA}
	\begin{align}\label{radial}
		0=\frac{1}{\Delta}\frac{\rd}{\rd r} \bigg(\Delta \frac{\rd R}{\rd r}\bigg)+\frac{1}{\Delta}\bigg[ \frac{1}{\Delta}\Big(\omega(r^2+a^2)-am\Big)^2-(\mu^2 r^2+\lambda)\bigg] R
	\end{align}
	and 
	\begin{align}\label{angular}
		0=\frac{1}{\sin\theta}\frac{\rd}{\rd\theta}\bigg( \sin\theta\frac{\rd S}{\rd \theta}\bigg)+\bigg[-\bigg( a \omega \sin\theta-\frac{m}{\sin\theta}\bigg)^2-\mu^2 a^2 \cos^2\theta+\lambda\bigg]S\,,
	\end{align}
respectively.

\subsubsection{Angular dependence}

Changing variable $z=\cos \theta$, the angular equation \eqref{angular} becomes 
\begin{align}\label{spherical eq}
	\frac{\rd}{\rd z}\bigg( (1-z^2)\frac{\rd S}{\rd z}\bigg)+\Big(  \Lambda_{\ell m} +g^2 (1-z^2)-\frac{m^2}{1-z^2}\Big)S=0 \, ,
\end{align}
with
\begin{align}
	\Lambda_{\ell m}(g)=\lambda_{\ell m}+2a\omega m-\mu^2 a^2\,,\qquad g^2=a^2 (\mu^2-\omega^2)=-a^2 \bar{k}^2 \,.
\end{align}
The labels $(\ell,m)$  correspond to successive solutions and eigenvalues $\Lambda_{\ell m}(g)$  to  \eqref{spherical eq}. Here $m$ is an integer such that $-\ell\leq m\leq \ell$. 

For $g^2=0$, i.e.~$\omega=\mu$ or $a=0$, \eqref{spherical eq} reduces to the associated Legendre equation, in which case $\Lambda_{\ell m}=\ell (\ell+1)$. The full angular dependence in such case is the spherical harmonics $Y_{\ell m}(\theta,\phi)\propto e^{i m\phi}P_\ell^{m}(\cos\theta)$, where $P_\ell^{m}(z)$ is the associated Legendre polynomial of the first kind.

The solutions to \eqref{spherical eq} for $g^2>0$ ($g^2<0$) are known as prolate (oblate) angular spheroidal wave functions, which we denote with $PS_{\ell m}(g,z)$, as in Mathematica \cite{reference.wolfram_2021_spheroidalps}. The basic properties of $PS_{\ell m}(g,z)$ can be found in e.g.~\cite{flammer_spheroidal_1957}. The parameter $g^2$ controls the deviation from $P_\ell^{m}(z)$, which has been taken to be zero throughout this paper. If $g^2$ is small but nonzero, we can include small corrections with the series expansions of $PS_{\ell m}(g,z)$ and $\Lambda_{\ell m}$ in powers of $g^2$, which to $O(g^2)$ read
\begin{multline}\label{small g PS}
	PS_{\ell m}(g,z)\\
	=P_\ell^m(z)+g^2 \left(\frac{(\ell-m+1) (\ell-m+2) P_{\ell+2}^m(z)}{2 (2\ell+1) (2\ell+3)^2}-\frac{(\ell+m-1) (\ell+m) P_{\ell-2}^m(z)}{2 (2\ell-1)^2 (2\ell+1)}\right)+O(g^4)
\end{multline}
and
\begin{align}
	\Lambda_{\ell m}(g^2) =\ell (\ell+1)-g^2\frac{2  \left(\ell^2+\ell+m^2-1\right)}{(2\ell-1) (2\ell+3)}+O(g^4)\,.
\end{align}
In this paper, for fixed $g$ we adopt the same normalization as in Mathematica \cite{reference.wolfram_2021_spheroidalps}:\footnote{For different $g$ and $g'$, $PS_{\ell,m}(g,z)$ and $PS_{\ell,m}(g',z)$ are not orthogonal. However, the non-orthogonality is small if both $g$ and $g'$ are small, as we have been assuming in this paper.}
\begin{align}\label{angular norm}
	\int_{-1}^1 dz PS_{\ell m}(g,z)PS_{\ell',m'}(g,z)=\delta_{\ell,\ell'}\delta_{m,m'}\frac{2(\ell+m)!}{(2\ell+1)(\ell-m)!}\,,
\end{align}
that is, we normalize $PS_{\ell m}(g,z)$ in the same way as $P_\ell^m(z)$. Therefore, our unit-normalized angular solution is
\begin{align}
	S_{\ell m}(\theta) = \sqrt{\frac{(2\ell+1)(\ell-m)!}{2(\ell+m)!}}PS_{\ell m}(g,\cos \theta) \approx \sqrt{\frac{(2\ell+1)(\ell-m)!}{2(\ell+m)!}} P_\ell^m(\cos\theta)+O(g^2)\,.
\end{align}
Clearly as $g\to 0$, $e^{im\phi}	S_{\ell m}(\theta)$ reduces to the usual spherical harmonics $Y_{\ell m}(\theta ,\phi)$.

\subsubsection{Radial dependence}

We rewrite \eqref{radial} as
\begin{align}
	0=\frac{\rd^2R}{\rd r^2}+\bigg(\frac{1}{r-r_+}+\frac{1}{r-r_-}\bigg) \frac{\rd R}{\rd r}+\frac{1}{\Delta}\bigg[ \frac{1}{\Delta}\Big(\omega(r^2+a^2)-am\Big)^2-(\mu^2 r^2+\lambda_{\ell m} )\bigg] R \, ,
\end{align}
which has singularities at $r=r_\pm$ and $r=\infty$. Making the change of variable
\begin{align}\label{change var}
	x=\frac{r-r_+}{r_- -r_+}
\end{align}
puts the equation into the form
\begin{align}\label{radial x}
	0=\frac{\rd R^2}{\rd x^2}+\bigg(\frac{1}{x}+\frac{1}{x-1} \bigg)\frac{\rd R}{\rd x}+\bigg(A^2_1 +\frac{A_2}{x}+\frac{A_3}{x-1} +\frac{A^2_4}{x^2}+\frac{A^2_5}{(x-1)^2} \bigg)R \, ,
\end{align}
where 
\begin{gather}
	A_1 =\bar{k}(r_+ -r_-)\,,\qquad A_2 =\frac{2 a^2 (m-2 a \omega )^2}{(r_+ -r_-)^2}- \left(\bar{k}^2+ \omega ^2\right) r_+^2+\lambda_{\ell m} \, , \nn\\
	A_3 =-\left[\frac{2 a^2 (m-2 a \omega )^2}{(r_+ -r_-)^2}- \left(\bar{k}^2+ \omega ^2\right) r_-^2+\lambda_{\ell m}\right] \, , \nn \\
	A_4 =\frac{ r_+  r_s \omega- m  a }{r_+ -r_-}\,,\qquad A_5 =\frac{r_-  r_s \omega- m  a }{r_+ -r_-} \,.
\end{gather}
Here we recall $\bar{k}$ is defined by $\omega^2 = \bar{k}^2+\mu^2$. Note that these expressions break down when the black hole is exactly extremal so that $r_+ =r_- $. Throughout this paper we focus on the case where the black hole is not exactly extremal.

To proceed, we introduce a new function $R(x)=e^{iA_1 x}(-x)^{i A_4}(1-x)^{i A_5}f(x)$ to bring the equation into the form
\begin{align}\label{Heun eq}
	f''(x)+\bigg( \alpha+\frac{1+\beta}{x}+\frac{1+\gamma}{x-1}\bigg)f'(x)+\bigg( \frac{C}{x}+\frac{D}{x-1}\bigg) f(x)=0
\end{align}
with
\begin{gather}
	\alpha = 2iA_1=2i\bar{k}(r_+ -r_-)\,,\qquad \eta=-A_2 =-\left[ \frac{2 a^2 (m-2 a \omega )^2}{(r_+ -r_-)^2}- \left(\bar{k}^2+ \omega ^2\right) r_+^2+\lambda_{\ell m} \right] \, ,  \nn\\
	\delta = A_3+A_2=-r_s \left(r_+-r_-\right) \left(\bar{k}^2+ \omega ^2\right)\,,\qquad \beta = 2i A_4=2i \frac{ r_+  r_s \omega- m  a }{r_+ -r_-}\, , \nn\\
	\gamma=2i A_5=2i \frac{ r_- r_s \omega- m  a }{r_+ -r_-} \, ,
\end{gather}
and
\begin{align}
	C=\frac{1}{2}-\frac{(1+\beta)(1+\gamma-\alpha)}{2}-\eta,\quad D=-\frac{1}{2}+\frac{(1+\beta+\alpha)(1+\gamma)}{2}+\delta+\eta\,.
\end{align}
The equation \eqref{Heun eq} is known as the confluent Heun equation,\footnote{Some properties of the confluent Heun equation and its solutions can be found in e.g.~\cite{ronveaux_heuns_1995}.} with linearly independent solutions
\begin{align}
	\text{HeunC}\left(\alpha,\beta,\gamma,\delta,\eta;x\right)\quad \text{ and }\quad (-x)^{-\beta}\text{HeunC}\left(\alpha,-\beta,\gamma,\delta,\eta;x\right)
\end{align}
normalized so that $\text{HeunC}\left(\alpha,\beta,\gamma,\delta,\eta;0\right)=1$. Therefore, we conclude that the full radial function is
\begin{align}\label{full radial x}
	R_{\omega\ell m}(x)=& \, e^{\frac{1}{2}\alpha x}(-x)^{\beta_m/2}(1-x)^{\gamma_m/2}\Big[ C_1 \text{HeunC}\left(\alpha,\beta_m,\gamma_m,\delta,\eta_{\ell m};x\right)\nn\\
	&+C_2 (-x)^{-\beta_m}\text{HeunC}\left(\alpha,-\beta_m,\gamma_m,\delta,\eta_{\ell m};x\right)\Big].
\end{align}

\subsubsection{Full solution and boundary condition at the horizon}

Putting everything together and restoring the original radial coordinate, the full solution is
\begin{align}\label{full sol}
	\Phi_{\omega\ell m}(t,r,\theta,\phi)=e^{-i\omega t}e^{i m\phi}S_{\ell m}(\theta) R_{\omega\ell m}(r)
\end{align}
with 
\begin{align}\label{full radial}
	R_{\omega\ell m}(r)=& \, \, e^{-i\bar{k}(r-r_+)}\left(-\frac{r-r_+}{r_- -r_+}\right)^{\frac{\beta_m}{2}}\left(-\frac{r-r_-}{r_- -r_+}\right)^{\frac{\gamma_m}{2}}\bigg[ C_1 \text{HeunC}\left(\alpha,\beta_m,\gamma_m,\delta,\eta_{\ell m};\frac{r-r_+}{r_- -r_+}\right)\nn\\
	&+C_2 \left(-\frac{r-r_+}{r_- -r_+}\right)^{-\beta_m} \text{HeunC}\left(\alpha,-\beta_m,\gamma_m,\delta,\eta_{\ell m};\frac{r-r_+}{r_- -r_+}\right)\bigg]\,.
\end{align}
Now we would like to pick the solution that is purely ingoing at the outer horizon $r=r_+$. This can be thought of as the solution that has a constant phase along an infalling null curve. As $r\to r_+$, the confluent Heun functions in \eqref{full radial} tend to one and the full solution approaches
\begin{equation}\label{near hor radial}
	\Phi_{\omega\ell m}(t,r\to r_+,\theta,\phi)
	=S_{\ell m}(\theta)\Big( C_1 e^{-i\omega (t-r^*)}e^{-i\omega r_+}e^{i m\phi_\text{out}}+C_2 e^{-i\omega (t+r^*)}e^{i\omega r_+}e^{i m\phi_\text{in}}\Big)  \,.
\end{equation}
Here we have introduced the infalling and outgoing Eddington-Finkelstein coordinates
\begin{align}\label{Change of var}
	t_\text{in} & = t +r^*\,,  	&\phi_\text{in} =\phi+ \frac{a}{r_+-r_-}  \ln \frac{r-r_+}{r-r_-} \, , \\
	t_\text{out} & = t -r^*\,, & \phi_\text{out} =\phi- \frac{a}{r_+-r_-}  \ln \frac{r-r_+}{r-r_-} \, ,
\end{align}
with the tortoise coordinate $r^*$ defined by
\begin{align}\label{tortoise}
	r^* =
	r+\frac{r_++r_-}{r_+-r_-} \left[ r_+ \ln \left( -\frac{r-r_+}{r_- -r_+}\right) -r_- \ln \left( -\frac{r-r_-}{r_- -r_+}\right) \right] \, .
\end{align}
Now, infalling null curves are those with constant $v=t+r^*$. Therefore, to impose the purely infalling boundary condition, we set $C_1=0$.

To summarize, the  solution for $\Phi$ with the correct infalling condition at the horizon is given by \eqref{full sol} with
\begin{equation}
\begin{split}
\label{ang rad sol}
	S_{\ell m}(\theta) &= \sqrt{\frac{(2\ell+1)(\ell-m)!}{2(\ell+m)!}}PS_{\ell m}(g,\cos \theta) \\
	R_{\omega\ell m}(r) &= |R_{\omega\ell m}(r_+) |  e^{-i\bar{k}(r-r_+)}\left( -\frac{r-r_+}{r_- -r_+}\right) ^{-\frac{\beta_m}{2}}\left( -\frac{r-r_-}{r_- -r_+}\right) ^{\frac{\gamma_m}{2}} 
	\\
&\qquad \times\text{HeunC}\left( \alpha,-\beta_m,\gamma_m,\delta,\eta_{\ell m};\frac{r-r_+}{r_- -r_+}\right)\,.
\end{split}
\end{equation}
Note that, far away from the black hole, the geometry is approximately flat and spatial gradients of the scalar field can be neglected, so that the field density for a single mode takes the form
\begin{align}
	\rho_{\mu\ell m} (r,\theta ,\phi)= -T^t{}_t\approx |\partial_t\Phi_{\mu\ell m} |^2+\mu^2 |\Phi_{\mu\ell m} |^2 =2\mu^2  |\Phi_{\mu\ell m} |^2 \, ,
\end{align}
where we are taking $\omega\approx \mu$. The field amplitude $|R_{\mu\ell m}(r_+) |$ at the horizon in \eqref{ang rad sol} is then related to the {\it angular average} $\bar\rho_{i,\ell m}$ of the field density at $r=r_i\gg r_s$ through
\begin{align}
	\bar\rho_{i,\ell m}&=2\mu^2  \int_{S^2}   |\Phi_{\mu\ell m}(r_i)|^2 =2\mu^2 |R_{\mu\ell m}(r_i)|^2\nn\\
	&=2\mu^2 |R_{\mu\ell m}(r_+)|^2\left|\text{HeunC}\left( \alpha,-\beta_m,\gamma_m,\delta,\eta_{\ell m};\frac{r_i-r_+}{r_- -r_+}\right)\right|^2\,.
\end{align}

\subsection{The $r\to \infty$ limit}

In this section we study the large distance behavior of the radial solution \eqref{ang rad sol}. To this end, we go back to the radial equation \eqref{radial x} and we write 
\begin{equation}\label{large x rewrite}
	R(x)= e^{\pm iA_1x} (1-x)^{-\frac{1}{2}}(-x)^{\mp i \frac{B}{2}} F(x) \, ,
\end{equation}
with 
\begin{equation}
 B =\sqrt{4 (A_3+ A_4^2+ A_5^2)-1} \,.
\end{equation}
The $\pm$ signs correspond to the  two linearly independent solutions. In terms of $F$ in \eqref{large x rewrite}, the radial equation \eqref{radial x} in the large-$x$ limit reads
\begin{equation}\label{1F1 eq}
	x F'' + (c_\pm \pm 2 i A_1 x )F' \pm 2 i A_1 a_\pm F= 0 \qquad (x\gg 0) \, ,
\end{equation}
where
\begin{equation}
	a_\pm= \pm\frac{A_2+A_3}{2 i A_1}+\frac{c_\pm}{2} \, , \qquad \qquad 
	c_\pm = 1\mp i B \,. 
\end{equation}
When $ \bar{k}^2 \neq 0$, the solution to \eqref{1F1 eq} is exactly the confluent hypergeometric function:
\begin{equation}
	 _1F_1(a_\pm,c_\pm,\mp i 2 A_1 x) \,. 
\end{equation}
The general large-$x$ radial solution is then a linear combination of the two $\pm$ solutions:
\begin{equation}\label{general large r}
	R (x) \approx \tilde{C_1} \, e^{i  A_1 x} (-x)^{\frac{-1-i B}{2}} \, _1F_1(a_+,c_+,-i 2 A_1 x)+ \tilde{C_2}\, e^{-i  A_1 x}  (-x)^{\frac{-1+iB}{2}} \, _1F_1(a_-,c_-,i 2 A_1 x)\; .
\end{equation}
Using the fact that
\begin{align}
	_1F_1(a,c,z\to\infty) \propto e^z z^{a-c} \left( 1+O\left( \frac{1}{z}\right) \right) \, ,
\end{align}
one has, for $\bar{k}^2 \neq 0$,
\begin{equation}
 	R(r)\approx C_3\, \frac{e^{i A_1r}}{r}e^{-i\frac{A_2+A_3}{4 A_1}\log \left( \frac{r-r_+}{r_+- r_- }\right) }+ C_4 \, \frac{e^{-iA_1r}}{r}e^{i\frac{A_2+A_3}{4 A_1}\log \left( \frac{r-r_+}{r_+- r_- }\right) } \,. 
\end{equation}
However, we are interested in the case $ \bar{k}^2 = 0$. In this limit the confluent hypergeometric functions in \eqref{general large r} become degenerate. 
We can  take $\bar k \to 0$ in \eqref{general large r} using
\begin{align}
	\lim_{\lambda\to0} \,_1F_1 \left(\frac{a}{\lambda},c;\lambda z\right)=\,_0F_1(c,az)\, .
\end{align}
Combining this with 
\begin{align}
	J_\alpha(x) =\frac{(\frac{x}{2})^2}{\Gamma(\alpha+1)} \,_0F_1\left( \alpha+1;-\frac{x^2}{4}\right) \,,
\end{align}
we have, for large $r$,
\begin{align}\label{large r gen}
	R (r)\; \stackrel{\bar{k}=0}{\approx} \; & \tilde{C_3}\, \frac{J_{B}\left(2\mu \sqrt{r_s \, r}\right)}{\sqrt{r}}+ \tilde{C_4}\, \frac{J_{-B}\left(2\mu \sqrt{r_s \, r}\right)}{\sqrt{r}}\, ,
\end{align}
where we have absorbed $r$-independent factors into the constants $\tilde{C_3}$ and $\tilde{C_4}$. For $\bar k =0$, the quantity $B$ is
\begin{align}
	B =2\sqrt{-\left( \ell+\frac{1}{2}\right) ^2+ \mu ^2 r_s (2r_s-r_+)} \, .
\end{align}
The precise relation between $\tilde{C_3}$, $\tilde{C_4}$ and the overall amplitude $|R_{\omega\ell m}(r_+) |$ in  \eqref{ang rad sol} depends on the scalar mass $\mu$, the black hole mass $r_s$ and spin $a$,
and the angular momentum quantum numbers $\ell$, $ m$. Such relation is usually not easy to find analytically in closed form.  In the following, we first discuss two limiting cases, i.e.~the particle and the wave regimes, for which it is possible to find a simple expression for the ratio  $|R(r_+)|^2/|R(r_i)|^2$ where $r_i\gg r_s$. We will later discuss the intermediate regime in Appendix~\ref{app:num}, where we will obtain an approximate connection formula by fitting the numerical solution of the radial equation. The results are summarized in Eq.~\eqref{eqn:guanhaos-fit}.

\paragraph{The particle limit.}

The approximation \eqref{large r gen} is valid when $\mu^2r_s(r-r_+)\gg 1$. For this to be valid all the way down to the near-horizon region $r \approx r_+$, we need in particular $\mu r_s \gg 1$. Now, if we further have 
\begin{align}
	 2\mu \sqrt{r_s(r-r_+)} \gg \left| B^2 \right| \, ,
\end{align}
we can use the asymptotic expression for the Bessel function
\begin{align}\label{bessel asym}
	J_\alpha (y) =\sqrt{\frac{2}{\pi y}}\left[  \cos\left( y-\frac{\alpha \pi}{2}-\frac{\pi}{4}\right) +O\left( y^{-1}\right) \right] \, , \quad y\gg \left| \alpha^2-\frac{1}{4} \right| 
\end{align}
to obtain a simple estimate. Excluding the extreme case $\mu r_s  \gtrsim \sqrt{r_i/r_s}$ and focusing on $\ell\sim O(1)$, we can use the following approximation for large enough $r$:
\begin{align}\label{far field large mass}
	|R_{\mu\ell m} (\to \infty)|^2\stackrel{\bar{k}=0}{\sim} r^{-\frac{3}{2}} \, , \quad \text{ for } \quad \mu r_s \gtrsim \frac{\ell +1}{2} \, .
\end{align}
We will later confirm numerically this approximation as well as its range of applicability. 

\paragraph{The wave (ultralight) limit.}

 For an ultralight scalar with mass $\mu^2\lesssim 1/r_s r_i$, the approximation \eqref{large r gen} does not hold for any distance $r\lesssim r_i$. However, to the leading order this case can be approximated in terms of a static massless scalar, i.e.~$\omega = \mu =0$. Writing
 \begin{align}
 	R(x)= \left( -\frac{x}{1-x}\right)^{\frac{i m a}{r_+ - r_-}}Y(x) \; ,
 \end{align}
equation \eqref{radial x} takes the hypergeometric form
\begin{align}
	x(1-x)  Y''(x)+\left(1+\frac{ 2 i m a}{r_+-r_-} -2x\right) Y'(x)+\ell (\ell+1) Y(x) =0 \, .
\end{align}
Picking the solution that is regular at the outer horizon, $r=r_+$, we have
\begin{align}
	R_{\ell m}(r)  =  |R_{\ell m}(r_+) | \left( \frac{r-r_+}{r-r_-}\right)^{\frac{i m a}{r_+ - r_-}} \, _2F_1\left(-\ell ,\ell+1 ;1+\frac{ 2 i m a}{r_+-r_-};\frac{r-r_+}{r_- - r_+}\right)  \, .
\end{align}
Finally, the asymptotics for the hypergeometric functions implies the following large-$r$ behavior: 
\begin{align}\label{far field ultralight}
	 |R_{\ell m}(r\to \infty) |^2 \propto r^{2\ell} \, , \qquad \text{ for }  \mu^2\lesssim 1/r_s r_i \, .
\end{align}

To conclude, we have derived the first and third line of \eqref{eqn:guanhaos-fit}. Unfortunately, the most interesting regime for the interplay between scalar hair and superradiance is the intermediate regime $2\sqrt{\rs/r_i}\ll\mu\rs\lesssim\frac12(\ell+1)$, i.e.~the second line of \eqref{eqn:guanhaos-fit}. In this case \eqref{far field ultralight} is not applicable, while \eqref{large r gen} is only valid down to some distance $r_0\gg r_+$ outside the near-horizon region such that $2\mu \sqrt{r_s r_0}\sim 1$. We will need to rely on numerical studies to obtain a good estimate, which is what we discuss next.

\subsection{Numerical results for hair solutions}
\label{app:num}

In this section we carry out numerical studies of the hair solution  in the different regimes, proving in particular evidence for  our estimates \eqref{eqn:guanhaos-fit}.

\subsubsection{$|R_+|/|R_i|$ as a function of $\alpha$ at $a_*=0$ and the three regimes}

To demonstrate the separation of three regimes \eqref{eqn:guanhaos-fit}, we first focus on the case with $a_*=0$. Recall that $m$ enters the radial function \eqref{ang rad sol} only through the combination $ma_*$, and thus a nonzero $m$ has no effect  in this case. Figure~\ref{ratio0} shows plots of the ratio $|R_+|/|R_i|$ as a function of $\alpha$ for $\ell=0,1,2$, with $R_+\equiv R(r_+)$ and $R_i \equiv R(r_i)$.\footnote{We will see in Figure~\ref{Rvsr} that $|R(r)|$ is an oscillatory  function of $r$ in the intermediate regime.
Therefore,  normalizing  the scalar profile $|R(r)|$  at fixed $r_i$ would result in the presence of spikes in $|R_+|/|R_i|$ as a function of $\alpha$, which correspond to the minima of the oscillations in Figure~\ref{Rvsr}.  
We get around this problem by sampling several $|R_+|/|R_i|$ values within the range $350r_s < r < 450r_s$, and take the minimum $|R_+|/|R_i|$ value. Some remnants of the spikes can still be seen in the plot in Figure~\ref{ratio0}, most prominently for $\ell=0$.}  Here we choose $r_i = 400r_s$ (we set $r_s=2$ when making the plots). 
\begin{figure}[H]
	\centering
	\includegraphics[scale=0.65]{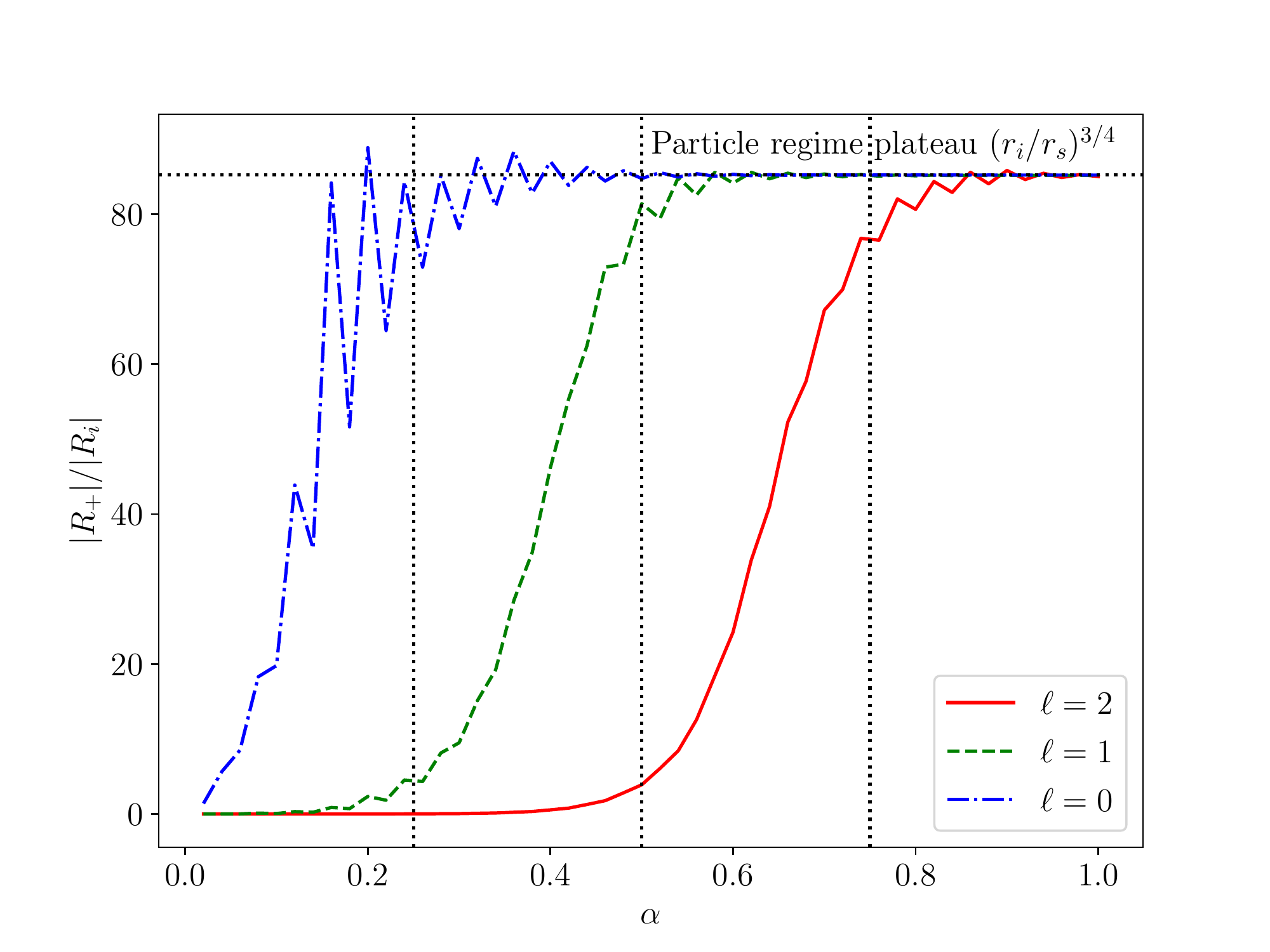}
	\caption{$|R_+|/|R_i|$ as function of $\alpha$, with $\ell=0,1,2$ at $a_*=0$. The horizontal black line marks the value of $|R_+|/|R_i|$ in the particle regime. The wave regime values behave as expected, although not distinguishable on this plot. The vertical dashed lines indicate the boundary $\alpha \simeq (\ell + 1)/2$, which was indicated in equation (\ref{eqn:guanhaos-fit}). }
	\label{ratio0}
\end{figure}
It is clear from Figure~\ref{ratio0} that, for  fixed $\ell$, there are three qualitatively different phases as $\alpha$ increases from 0 to 1: two asymptotic flattened regions and an intermediate phase. The flattening of the ratio $|R_+|/|R_i|$ as $\alpha\to 1$ and $\alpha \to 0$ correspond respectively to the ``particle" and ``wave"   regimes studied analytically in the previous section.\footnote{Note that the upper bound of the wave regime here is the boundary between ``regime II" and ``regime III" defined in \cite{Hui:2019aqm}, not the one between ``regime I" and ``regime II". However, the behavior of $|R_+|/|R_i|$  is identical in regime I and regime II, which means we cannot distinguish them by plotting $|R_+|/|R_i|$. We therefore merge regime I and II of \cite{Hui:2019aqm} into  a single one in our discussion, and call it the ``wave regime". }
As expected, in the particle regime, the value of $|R_+|/|R_i|$ plateaus around $(r_i/r_s)^{3/4}$.\footnote{We have chosen $r_i = 400 r_s$ in our numerical calculations, but the height of the plateau in Figure~\ref{ratio0} is actually $350^{3/4}$. This is an artifact of our procedure of smoothing out the oscillations in the intermediate regime. The behavior of $|R_+|/|R_i|$ remains qualitatively the same.} The numerical values in the wave regime agree with the $(r_i/r_s)^{-2\ell}$ approximation, although not distinguishable in Figure~\ref{ratio0} due to their small size. Figure~\ref{Rvsr} illustrates the typical behaviors for the radial function $|R(r)|^2/|R_+|^2$ as a function of $r$ in the three regimes: monotonically increasing (wave regime), oscillatory (intermediate regime), and monotonically decreasing (particle regime).
\begin{figure}[H]
	\centering
	\includegraphics[scale=0.65]{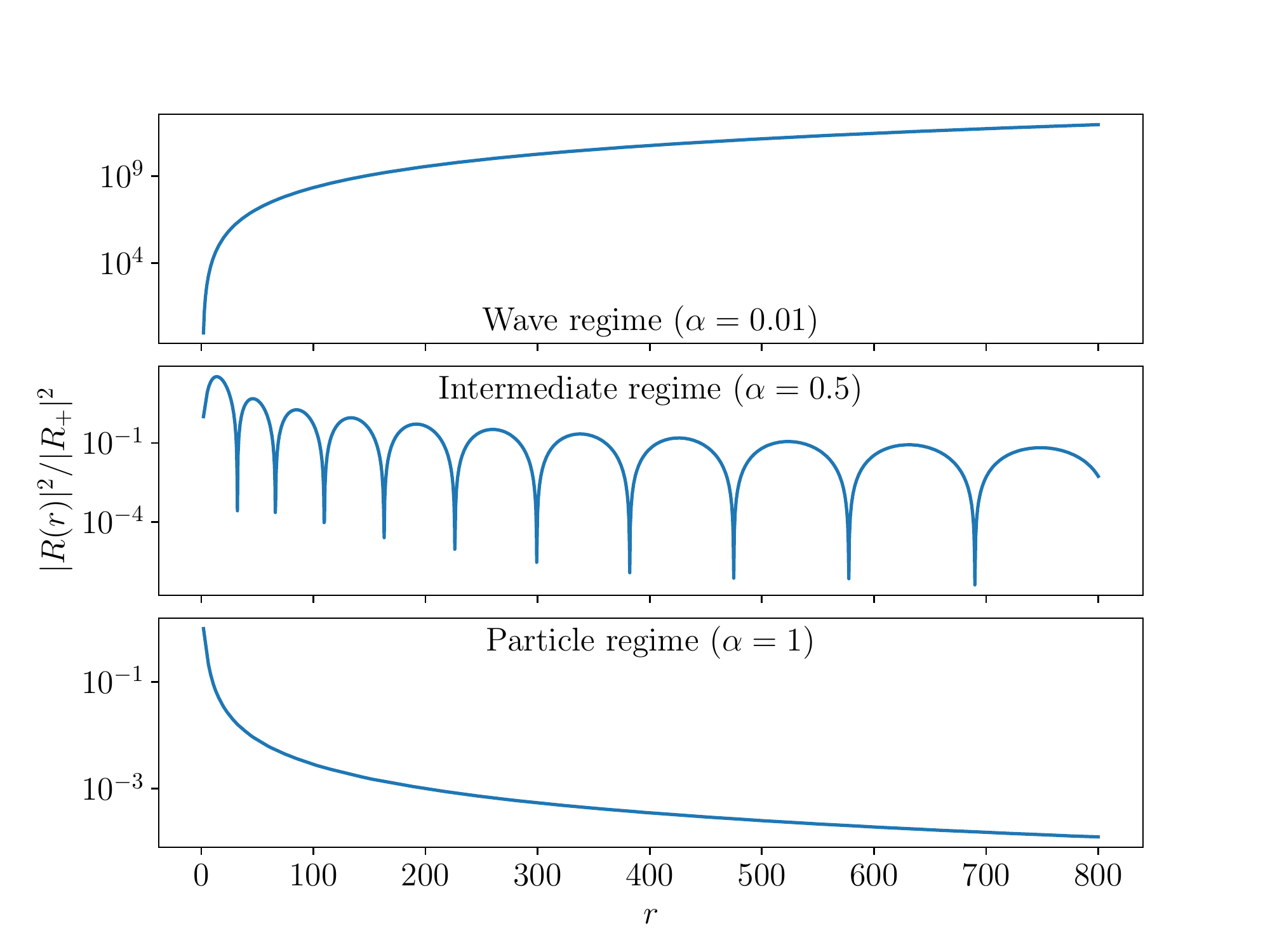}
	\caption{$|R(r)|^2/|R_+|^2$ as function of $r$ when $a_*=0$ and $(\ell, m) = (2,2)$ in the three regimes. }
	\label{Rvsr}
\end{figure}
In the previous section, we  obtained  the analytic approximations  \eqref{far field large mass} and \eqref{far field ultralight} for the large-$r$ behavior of   $R(r)$, and thus for the ratio $|R_+|/|R_i|$, in the particle and wave limits. Even though we have little analytic control over the intermediate regime, we can use the numerical results in Figure~\ref{ratio0} to extract some  simple  estimates for the scaling of $R(r)$. The results are   summarized in Eqs.~\eqref{eqn:guanhaos-fit}. In Figure~\ref{NumVsAppr}, we show that a very good agreement between \eqref{eqn:guanhaos-fit} and the exact numerical results is achieved within $O(1)$ error for $\ell=1$ and $\ell=2$. 
\begin{figure}[H]
	\centering
	\includegraphics[scale=0.65]{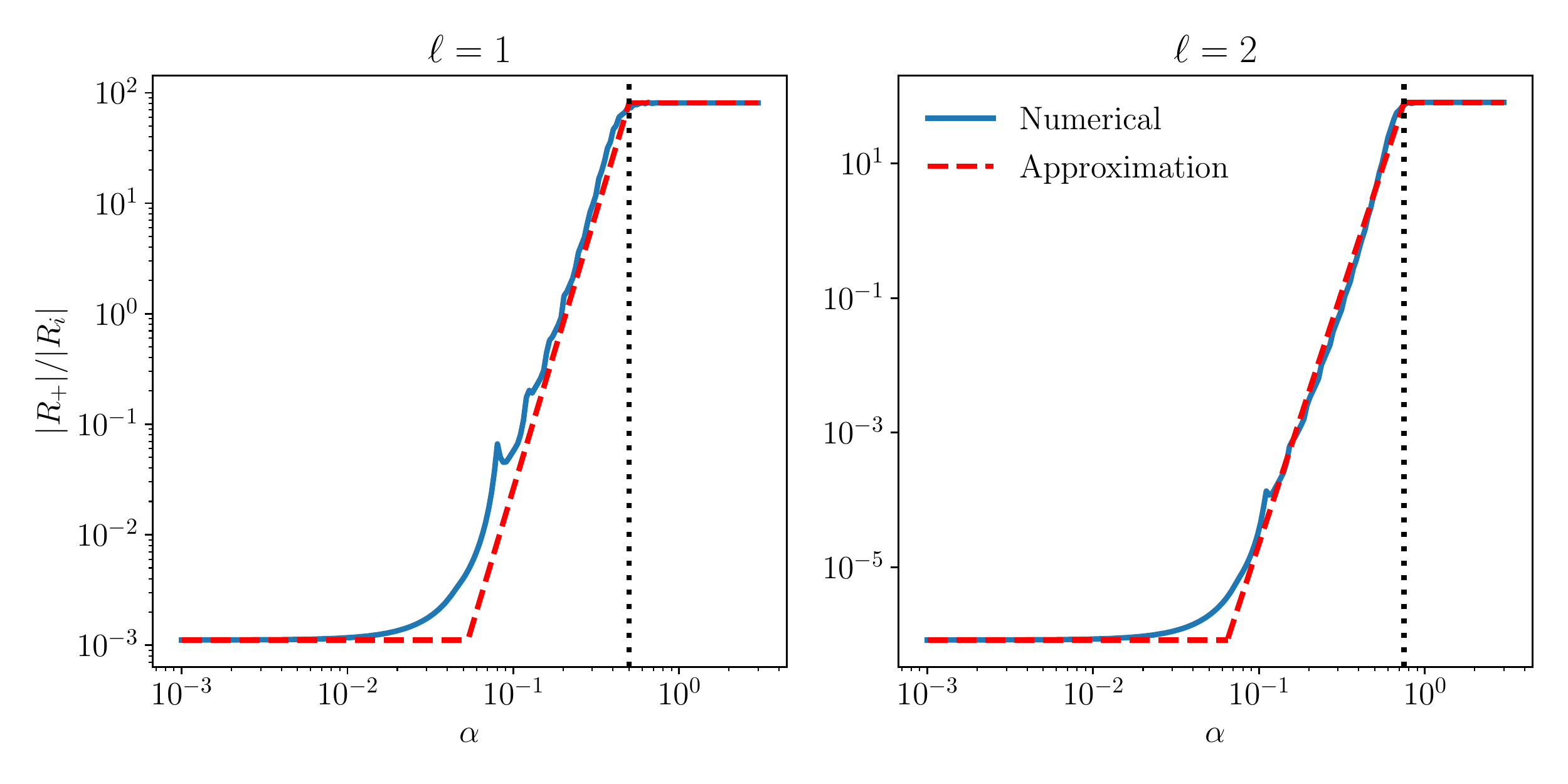}
	\caption{Estimates (red dash) vs.~exact numerical results (blue solid) for $|R_+|/|R_i|$, with $\ell = 1$ and $\ell = 2$ for $a_*=0$.}
	\label{NumVsAppr}
\end{figure}
The region where our approximation is the least accurate is the transition between the wave and intermediate regimes, which we simply define as the point where the approximations \eqref{eqn:guanhaos-fit} for the two regimes meet. It is worth noting that numerical studies show that this bound decreases with increasing the ratio  $r_i/r_s$, as described in our approximation, while the rate of the drop in the intermediate regime and the lower bound of the particle regime are not affected by this ratio. Therefore, for a larger $r_i$, $|R_+|/|R_i|$  is smaller in the wave regime due to a longer drop.

\subsubsection{$|R_+|/|R_i|$ on the Regge plane}

In the previous section we have identified   three different  regimes for the study of the scalar hair solution around a non-rotating Schwarzschild black hole  ($a_*=0$). Here we study the effect of turning on $a_*$. The main conclusion is that our approximations \eqref{eqn:guanhaos-fit} receive modifications of at most $O(1)$ unless $a_*>0.95$. As a first example, Figure~\ref{ratio1} shows the ratio $|R_+|/|R_i|$ as a function of $\alpha$ at $a_* \simeq 0.505$. 
\begin{figure}[H]
	\centering
	\includegraphics[scale=.65]{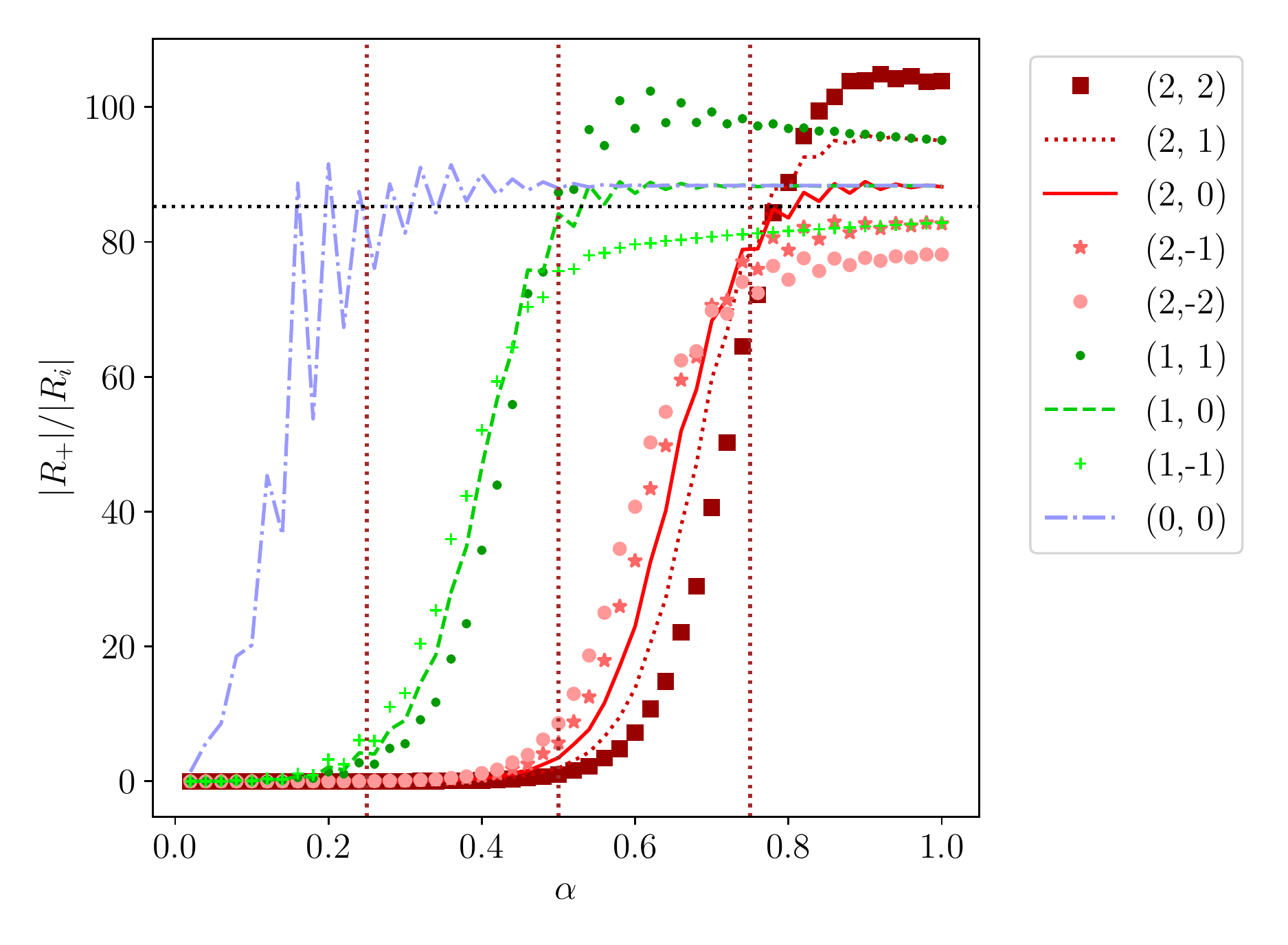}
	\caption{$|R_+|/|R_i|$  as function of $\alpha$ for $a_* \simeq 0.505$, with low values of $(\ell,m)$. The horizontal and vertical dotted lines are the same as in Figure \ref{ratio0}. Note that this figure deviates from figure \ref{ratio0} by only a small fraction.}
	\label{ratio1}
\end{figure}
Compared with Figure \ref{ratio0}, $|R_+|/|R_i|$ for the same $\ell$ but different $m$ now behave differently. We see that the value of $|R_+|/|R_i|$ in the particle regime now depends on the value of $m$, while it remains largely unaffected in the wave regime. Also, the boundaries separating the regimes are shifted, depending on the sign and size of $ma_*$. However, the effect of a nonzero $a_*$ on these plots is within $O(1)$ unless $a_* \gtrsim 0.95$. The dependence of $|R_+|/|R_i|$ on $a_*$ in the three different values of $\alpha$ is illustrated in Figure \ref{ratioastar}.

\begin{figure}[H]
	\centering
	\includegraphics[width=1\textwidth]{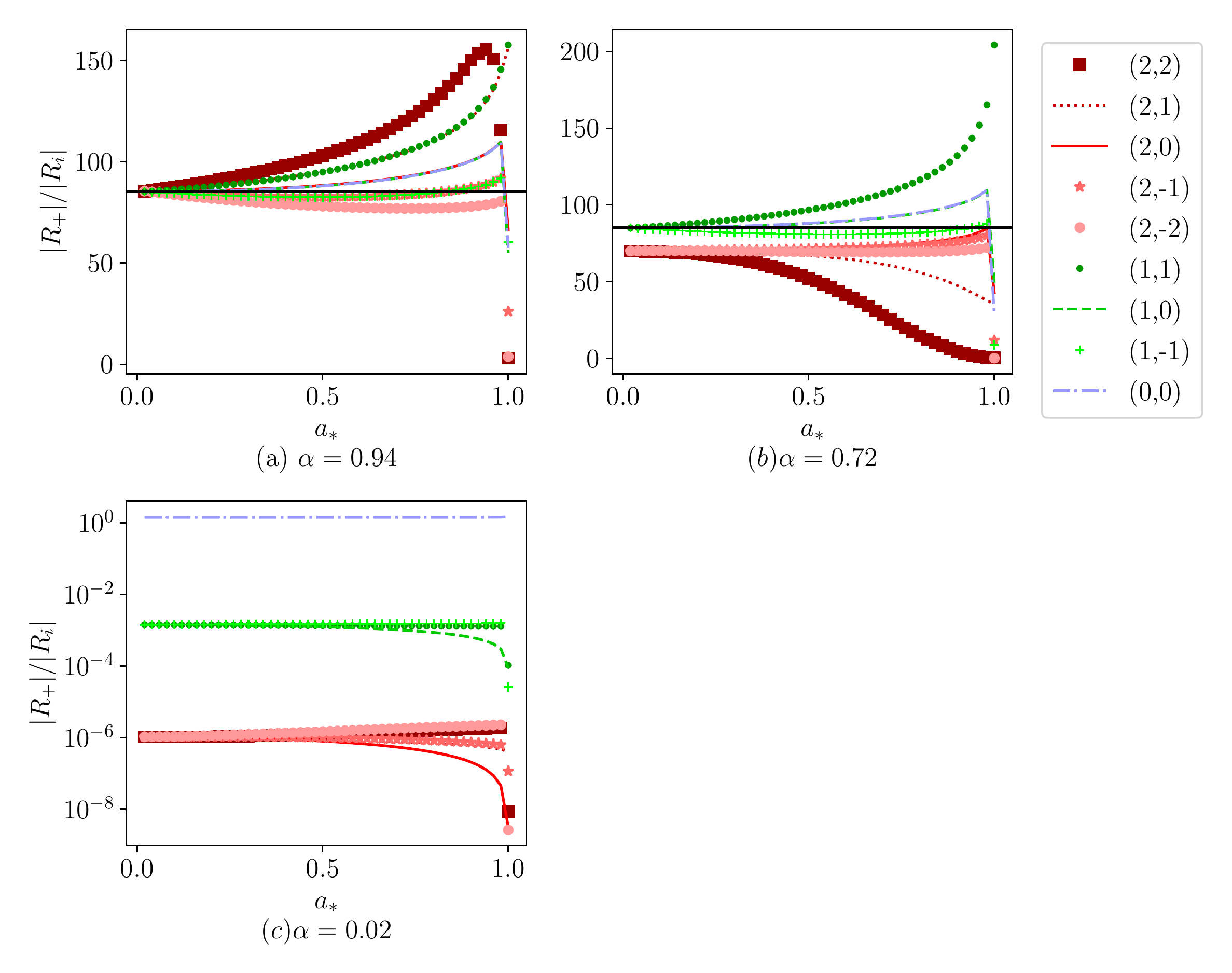}
	\caption{$|R_+|/|R_i|$  as function of $a_*$ for low values of $(\ell,m)$  for $\alpha=0.02,0.72$ and 0.94.}
	\label{ratioastar}
\end{figure}
Finally, Figure \ref{CRegge} shows the value of $|R_+|/|R_i|$ on the Regge plane $(\alpha,a_*)$, for $(\ell,m) = (2,2)$. In this figure, the deep blue and light yellow regions correspond to the wave and particle regimes respectively, while the greenish region is the intermediate regime. 
\begin{figure}[H]
	\centering
	\includegraphics[width=0.65\textwidth]{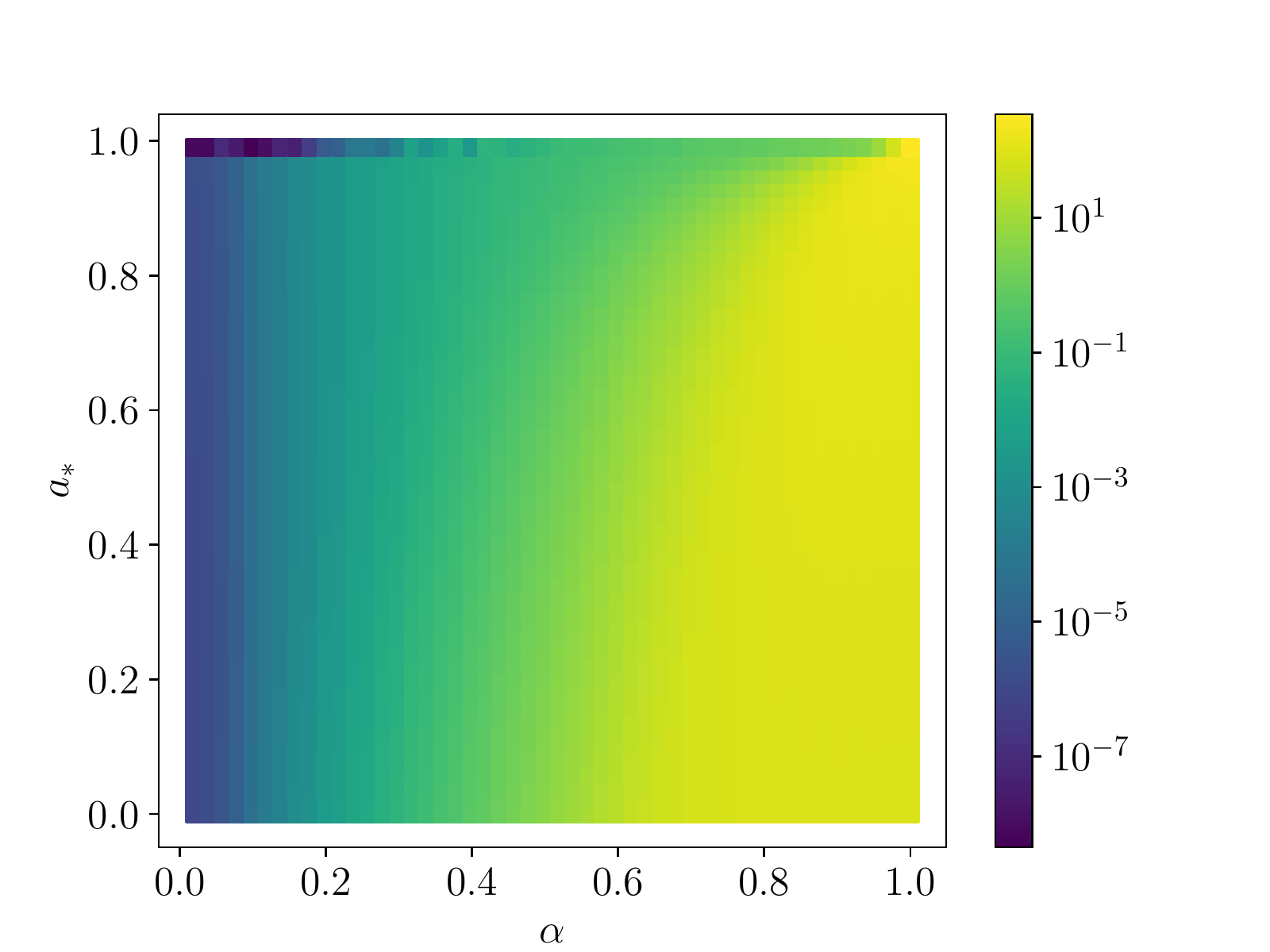}
	\caption{$|R_+|/|R_i|$ on the Regge plane $(\alpha,a_*)$ for $(\ell,m) = (2,2)$.}
	\label{CRegge}
\end{figure}
From both Figure \ref{ratioastar} and \ref{CRegge}, it is clear that the change of transition points between the three regimes as $a_*$ increases is well within $O(1)$ unless the black hole is near-extremal ($a_*>0.95$). Therefore, we have established the validity of the approximations \eqref{eqn:guanhaos-fit} up to $a_* \sim 0.95$, and therefore justifying dropping the $a_*$ (and thus $m$) dependence in \eqref{eqn:guanhaos-fit}. When the black hole is near extremal ($a_*>0.95$), all our analysis breaks down and a separate discussion is required.

\section{Superradiance}
\label{app2}

In this appendix, we will review some aspects of black hole superradiance and the corresponding system of the ``gravitational atom''.

\subsection{Bound states}

At distances much larger than the Schwarzschild radius, $r \gg \rs$, it is convenient to consider the following ansatz for the scalar field $\Phi$:
\beq
\Phi(t,\mathbf r)=\frac{1}{\sqrt{2\mu}}\left[\psi(t,\mathbf r)e^{-i\mu t}+\psi(t,\mathbf r)^*e^{i\mu t}\right] ,\eeq
where $\psi$ is a complex scalar field which varies on a timescale longer than $\mu^{-1}$. It can be shown that, to leading order in an expansion in powers of $\rs/r$, the Klein-Gordon equation $(\nabla^\nu\nabla_\nu-\mu^2)\Phi=0$  reduces to
\begin{equation}
i\frac{\partial\psi}{\partial t} = \biggl(-\frac1{2\mu}\nabla^2-\frac\alpha{r}\biggr)\psi\,,\qquad\text{where }\, \alpha \equiv \frac{\mu\rs}2\,,
\label{eqn:Schrodinger}
\end{equation}
which is  equivalent to the Schr\"odinger equation for the hydrogen atom, if we identify $\alpha$ with the fine structure constant. When $\psi$ is taken to (exponentially) vanish at infinity, Eq.~\eqref{eqn:Schrodinger} is then  solved by hydrogenic-like discrete bound states, whose spectrum is the familiar
\begin{equation}
\omega_{n\ell m}= \mu\biggl(1 -\frac{\alpha^2}{2n^2}\biggr).
\end{equation}
Higher-order corrections in powers of $\alpha$ will be present, due to
\begin{enumerate}
\item higher-order terms that we neglected in \eqref{eqn:Schrodinger};
\item the causal ingoing boundary conditions at the horizon, which differ from the demand of regularity at the origin of the hydrogen atom.
\end{enumerate}
The corrections of second type are particularly relevant, because they introduce a small imaginary part to $\omega_{n\ell m}$, making the population of the bound states either exponentially decrease or increase over time. The first terms in expansions are \cite{Baumann:2019eav}
\begin{align}
\label{eqn:ReOmega}
\Re(\omega_{n\ell m})&=\mu\biggl(1-\frac{\alpha^2}{2n^2}-\frac{\alpha^4}{8n^4}+f_{n\ell}\frac{\alpha^4}{n^3}+h_\ell\frac{ma}\rs\frac{\alpha^5}{n^3}+\ldots\biggr),\\
\label{eqn:ImOmega}
\Im(\omega_{n\ell m})&=4\frac{r_+}\rs C_{n\ell}g_{\ell m}\bigl(m\Omega_+-\Re(\omega_{n\ell m})\bigr)\alpha^{4\ell+5},
\end{align}
where the expressions of the coefficients $f_{n\ell}$,
$h_\ell$,$C_{n\ell}$ and $g_{\ell m}$ are given in
\cite{Baumann:2019eav}. The most relevant feature, to our purposes, is
that $\Im(\omega_{n\ell m})$ changes sign in correspondence of the
superradiance threshold, $m\Omega_+=\Re(\omega_{n\ell
  m})\approx\mu$. We thus see that the states are
\textit{quasi}-bound, as some of them decay, while others grow by superradiance.

\subsection{Fluxes and nonlinear evolution}
\label{app:superradiant-nonlinear}

Our analysis  has so far only dealt with the linear regime; the superradiant states, however, will eventually extract enough mass and angular momentum from the black hole to significantly change its parameters. To study this phase of the evolution, we can write down the fluxes of energy and angular momentum of the scalar field, under the assumption that only one $(n,\ell,m)$ mode is present:
\begin{align}
T^r{}_t&=g^{rr}(\partial_r\Phi^*\partial_t\Phi+\partial_t\Phi^*\partial_r\Phi)=2\frac\Delta{\varrho^2}\Im(\omega R'^*R)|S|^2e^{2\Im(\omega)t},\\
T^r{}_\phi&=g^{rr}(\partial_r\Phi^*\partial_\phi\Phi+\partial_\phi\Phi^*\partial_r\Phi)=-2\frac\Delta{\varrho^2}m\Im(R'^*R)|S|^2e^{2\Im(\omega)t}.
\end{align}
From the near-horizon limit of the radial part of the Klein-Gordon equation,
\begin{equation}
\label{eqn:near-horizon-radial2}
\Delta\frac{\rd}{\rd r}\biggl(\Delta\frac{\rd R}{\rd r}\biggr)+\rs^2r_+^2(\omega-m\Omega_+)^2R=0,
\end{equation}
we can extract the near-horizon behavior of $R(r)$ and write
\begin{align}
\label{eqn:T^r_t2}
T^r{}_t(r_+)&=2\frac{r_sr_+}{\varrho^2}(|\omega|^2-\Re(\omega)m\Omega_+)\Phi^*\Phi(r_+),\\
T^r{}_\phi(r_+)&=-2m\frac{r_sr_+}{\varrho^2}(\Re(\omega)-m\Omega_+)\Phi^*\Phi(r_+).
\label{eqn:T^r_phi2}
\end{align}
Equating the fluxes of mass and angular momentum to the change in the parameters of the black hole, and performing an angular integral, we arrive to
\begin{align}
\label{eqn:mass-evolution}
\frac{\rd \rs}{\rd t}&=4G\sum_{n,\ell,m}\rs r_+(|\omega_{n\ell m}|^2-\Re(\omega_{n\ell m})m\Omega_+)|R_{n\ell m}(r_+)|^2,\\
\label{eqn:J-evolution}
\frac{\rd(a\rs)}{\rd t}&=4G\sum_{n,\ell,m}\rs r_+m(\Re(\omega_{n\ell m})-m\Omega_+)|R_{n\ell m}(r_+)|^2.
\end{align}
In these equations, we are summing over all $(n,\ell,m)$ modes. Technically, this would not be allowed, because $T_{\mu\nu}$ is quadratic in the field and would thus contain interference terms. However, in the limit of small spheroidicity ($a^2(\mu^2-\omega^2)\approx a^2\mu^2\alpha^2/(2n^2)\ll1$), the angular integral kills the interference terms among states with different $(\ell,m)$ because of the orthonormality of spherical harmonics. Interferences between overtones with same angular momentum, instead, oscillate with frequency $\omega_n-\omega_{n'}\approx(1/2)\mu\alpha^2(1/n'^2-1/n^2)$. Comparing the power of $\alpha$ with \eqref{eqn:ImOmega}, it is easy to see that this frequency is much faster than the superradiance growth timescale, therefore it is same to mediate these interferences to zero.

The way equations (\ref{eqn:mass-evolution}) and (\ref{eqn:J-evolution}) are used to describe the nonlinear evolution of a superradiance-generated cloud has been described in Section \ref{sec:fluxes-evolution}.

\subsection{Superradiance and area law}
\label{sec:superradiance-area}

The area of the horizon of a Kerr black hole is $4\pi\rs r_+$. Using that
\begin{equation}
\label{eqn:derivative-rirr}
\frac{\rd(\sqrt{\rs r_+})}{\rd t}=\frac{\sqrt{\rs r_+}}{2r_+-\rs}\biggl(\frac{\rd\rs}{\rd t}-\frac{a}{\rs r_+}\frac{\rd(a\rs)}{\rd t}\biggr),
\end{equation}
we can combine \eqref{eqn:mass-evolution} and \eqref{eqn:J-evolution} to get
\begin{equation}
\label{eqn:rirr-evolution}
\frac{\rd(\sqrt{\rs r_+})}{\rd t}=4G\sum_{n,\ell,m}\frac{(\rs r_+)^{3/2}}{2r_+-\rs}|\omega_{n\ell m}-m\Omega_+|^2|R_{n\ell m}(r_+)|^2\ge0.
\end{equation}
This shows that the second law of black hole thermodynamics,
$\rd(\text{Area})/\rd t\ge0$, is respected. Moreover, we see that
along the trajectory due to superradiance, we have
\begin{equation}
\frac{\rd(\sqrt{\rs r_+})}{\rd\rs}=\frac{\sqrt{\rs r_+}}{(2r_+-\rs)\mu}\bigl(\mu-m\Omega_+\bigr),
\end{equation}
where we used (\ref{eqn:nonlin-1-mode-super}) and (\ref{eqn:derivative-rirr}). Using $\rs$ to parametrize the curve of the superradiance trajectory, this equation is telling us that the derivative of the area along the said curve vanishes at $\mu=m\Omega_+$. The superradiance trajectories are thus tangent, on the threshold, to the constant area lines, see Figure \ref{fig:regge-introductory}. This means that the evolution is a quasi-adiabatic process in the vicinity of the superradiance threshold.

\section{A toy model}
\label{sec:toy-model}

Equations \eqref{eqn:M:acc+superr}, \eqref{eqn:J:acc+superr} and \eqref{eqn:R-Im-omega} contain some complications that may hide the physically relevant parts. Consider the following system of differential equations:
\begin{align}
\label{eqn:x-prime}
X'&=\lambda_X-(Y-X)Z,\\
\label{eqn:y-prime}
Y'&=\lambda_Y-2(Y-X)Z,\\
\label{eqn:z-prime}
Z'&=(Y-X)Z.
\end{align}
Here, the variables $X$ and $Y$ play the role of the two coordinates in the Regge plane, say $\alpha$ and $a_*$, while the variable $Z$ plays the role of the mass of the cloud, say $|R(r_+)|^2$. The line $Y=X$ is taken here to represent the superradiance threshold, $\mu=m\Omega_+$, and we used the fact that the growth rate ($\Im(\omega)\propto(m\Omega_+-\mu)$, see \eqref{eqn:ImOmega}) is proportional to the distance from the threshold. Finally, the parameters $\lambda_X$ and $\lambda_Y$ mock up the accretion rates of mass and angular momentum. The factor of 2 in \eqref{eqn:y-prime} is there to make sure that the slope of the threshold is smaller than the slope of the superradiance flux (any other larger-than-1 number would work equally well).

It is easy to see that equations \eqref{eqn:x-prime}, \eqref{eqn:y-prime} and \eqref{eqn:z-prime} imply $X+Z=\lambda_Xt+C_X$ and $Y+2Z=\lambda_Yt+C_X$, where $C_X$ and $C_Y$ are integration constants, and thus $Y-X=(\lambda_Y-\lambda_X)t+C_Y-C_X-Z$. Plugging this back into \eqref{eqn:z-prime}, we get
\begin{equation}
Z'=\bigl((\lambda_Y-\lambda_X)t+C_Y-C_X-Z\bigr)Z.
\end{equation}
Recall that the variable $Z$ represents the mass of the cloud: we thus require $Z(0)>0$. This implies that $Z(t)>0$ for every $t$, as otherwise $Z(t)$ would cross the trivial solution $Z(t)=0$. If $\lambda_Y>\lambda_X$, the line $Z(t)=(\lambda_Y-\lambda_X)t+(C_Y-C_X)$ is an attractor for all $t>0$: at large times, the solution can be expanded perturbatively as
\begin{equation}
Z(t)=(\lambda_Y-\lambda_X)t+(C_Y-C_X)-\text{const.}\times\exp\biggl(-(\lambda_Y-\lambda_X)\frac{t^2}2-(C_Y-C_X)t\biggr)+\ldots
\end{equation}
In this case, the mass of the cloud increases (linearly) with time. If $\lambda_Y<\lambda_X$, instead, the line $Z(t)=(\lambda_Y-\lambda_X)t+(C_Y-C_X)$ will only be an attractor for a finite time, during which the mass of the cloud will decrease linearly; afterwards, the line $Z(t)=0$ will become the new attractor: the solution at large times will be
\begin{equation}
Z(t)=\text{const.}\times\exp\biggl((\lambda_Y-\lambda_X)\frac{t^2}2+(C_Y-C_X)t\biggr)+\ldots
\end{equation}
The two cases $\lambda_Y>\lambda_X$ and $\lambda_Y<\lambda_X$ are in obvious correspondence with the over-superradiance and under-superradiance we described in Section \ref{sec:accretion+superradiance}.

In both cases, when $Z(t)$ is attracted to the line $(\lambda_Y-\lambda_X)t+(C_Y-C_X)$, we see that
\begin{equation}
Y-X=(\lambda_Y-\lambda_X)t+C_Y-C_X-Z
\end{equation}
is attracted to zero. The system therefore drifts along the threshold $Y=X$ as long as the mass of the cloud, $Z$, is large enough. From
\begin{equation}
Y-2X=(\lambda_Y-2\lambda_X)t+C_Y-2C_X,
\end{equation}
we can find the approximate evolution of the individual coordinates $X$ and $Y$, using $Y-2X\approx -X\approx -Y$. Of course, in this toy model we have treated the parameters $\lambda_X$ and $\lambda_Y$ as free. In the realistic case, if the second law of black hole thermodynamics holds (which requires the null energy condition and global hyperbolicity), the possible accretion fluxes are constrained to those that increase the black hole area. We saw in Section (\ref{sec:superradiance-area}) that the superradiance trajectory is tangent, on the threshold, to the constant-area lines. As in this toy model the superradiance trajectory is $Y=2X$, the said constraint on accretion would mean $\lambda_Y<2\lambda_X$.

\bibliographystyle{utphys}
\addcontentsline{toc}{section}{References}
\bibliography{BHbib}

\end{document}